%% file: Wigner_in_relativistic_QFT_All.tex
\documentclass[prd,tightenlines,nofootinbib,superscriptaddress,twocolumn]{revtex4}

\input{Wigner_in_relativistic_QFT_common}

\begin{document}

\input{Wigner_in_relativistic_QFT_body}
\bibliographystyle{bib-style}
\bibliography{biblio}

\appendix 

\widetext

\input{Wigner_in_relativistic_QFT_app}

\end{document}

%% file: Wigner_in_relativistic_QFT_common.tex
\usepackage[colorlinks=true,linkcolor=black]{hyperref}
\usepackage{graphicx}
\usepackage[utf8]{inputenc}
\usepackage[T2A,T1]{fontenc}	
\usepackage{textcomp}
\usepackage{amsmath}
\usepackage{amssymb}
\usepackage{amsthm}
\usepackage{amsfonts}
\usepackage{booktabs}
\usepackage{braket}
\usepackage{bbm}
\usepackage{xr-hyper}
\usepackage[capitalise]{cleveref}
\usepackage{color}
\usepackage{dsfont}
\usepackage[dvipsnames]{xcolor}
\usepackage{esint}
\usepackage{mathrsfs}
\usepackage{multirow}	
\usepackage{refcount}
\usepackage{slashed}
\usepackage{siunitx}
\usepackage{tabularx}
\usepackage{pifont}
\usepackage{xcolor}

\usepackage{todonotes}
\crefformat{section}{Sec.~#2#1#3}
\crefformat{subsection}{Sec.~#2#1#3}
\crefformat{subsubsection}{Sec.~#2#1#3}

\usepackage{tikz}
\usepackage{circuitikz}
\usetikzlibrary{shapes.geometric,calc,intersections,
decorations.markings,decorations.text,decorations,patterns, arrows,
decorations.pathmorphing,fpu,math,babel}

\newcommand{\me}{\mathrm{e}}
\newcommand{\mi}{\mathrm{i}}
\newcommand{\md}{\mathrm{d}}

\DeclareMathOperator{\trace}{tr}

\DeclareMathOperator{\arcosh}{arcosh}


%% file: Wigner_in_relativistic_QFT_body.tex
\title{A time-frequency approach to relativistic correlations in quantum field theory}

\author{{\bf Benjamin Roussel}}\email{benjamin.roussel@esa.int}
\author{{\bf Alexandre Feller}}\email{alexandre.feller@esa.int}
\affiliation{Advanced Concepts Team, European Space Agency, Noordwijk, 2201 AZ, Netherlands}

\begin{abstract}
	Moving detectors in relativistic quantum field theories reveal the 
	fundamental entangled structure of the vacuum which manifests, 
	for instance, through its thermal character when probed by a uniformly 
	accelerated detector. In this paper, 
	we propose a general formalism inspired both from signal processing
	and correlation functions of quantum optics to analyze the response of 
	point-like detectors following a generic, non-stationary trajectory.
	In this context, the Wigner representation of the first-order correlation
	of the quantum field is a natural time-frequency tool to understand 
	single-detection events. This framework offers a synthetic perspective
	on the problem of detection in relativistic theory and allows us to 
	analyze various non-stationary situations (adiabatic, periodic) and how
	excitations and superpositions are deformed by motion. It opens up 
	interesting perspective on the issue of the definition of particles.
\end{abstract}

\maketitle

\section{Introduction} 

One of the fundamental differences between relativistic quantum field
theories and quantum mechanics is the deeply entangled structure of 
quantum fields. While this can be understood in a general formal setting
\cite{Hollands-2018,Witten-2018}, one of the
clearest phenomena illustrating this is the entangled structure
of the vacuum state which is revealed by its thermal character in curved
spacetime \cite{Hawking_1975} or by a uniformly accelerated observer
\cite{Unruh-1976}, respectively known as the Hawking and Unruh effects.

The correlated nature of the vacuum is nicely probed by considering
a moving detector in spacetime coupled to the quantum field.
Such models are known as Unruh-Dewitt detectors. The thermal 
nature of the vacuum is then seen through its photo-detection 
response. Many questions can then be addressed such as the role of
causality \cite{Schlicht-2004}, the behavior under different motions
\cite{Grove-1983, Padmanabhan-2010, Doukas-2013}
or the effect of the switching function of the detector
\cite{Satz-2007,Fewster-2016}. 
A similar photo-detection approach has been used in quantum
optic since the work of Glauber on coherence
functions~\cite{Glauber-1963-1,Glauber-1963-2} and has been extended
to condensed matter situations~\cite{Bocquillon-2014}.

However, the interpretation of these responses in the context
of relativistic quantum field theory in flat or in curved spacetime 
is subtler than in quantum optics since no general notion of particles
can be defined in the standard way. The qualitative reason comes 
from the non existence of a global definition of time.
Two directions can then be taken. The first direction is to have an 
operational perspective: particles are defined through the response 
signal of the detector itself \cite{Unruh-1976}. 
The second direction follows a more 
pragmatic interpretation of the detector's response: the detector
is simply seen as a ``fluctuometer'', as a system that responds to 
the fluctuations of the quantum field. The signal should not 
\emph{a priori} be interpreted as coming from a particle content
\cite{Sciama-1981,Grove-1983,Takagi-1986}. 

Following this latter point of view is similar to adopt a signal processing
perspective, which we will adopt here. In a non stationary
context, physically meaningful information can be extracted from the
signal by performing a time-frequency analysis, giving us access to 
the evolution in time of the frequency content of the response.
Time-frequency (or time-scale) analysis is now a major tool in signal
processing, especially the Wigner function distribution~\cite{Flandrin-1998}.
Historically, the Wigner function has been introduced in quantum
mechanics as a phase space representation of the quantum state
\cite{Wigner-1932}. This distribution is now widely used in quantum
optics~\cite{Haroche-2006} and has been recently adapted to analyze
coherence properties of electron in the quantum Hall regime
\cite{Ferraro-2013, Roussel-2017}.

In this paper, we present a unified view on the response of 
a moving detector probing a relativistic quantum field using a 
time-frequency approach of the correlation functions of the 
field based on the Wigner distribution.
The main goal is to introduce the a good framework to analyze the 
response of a detector in general situations where the state of the
field can contain excitations, for an arbitrary trajectory.
This is achieved by using  both the correlation functions formalism 
and a time-frequency analysis. Physically realistic situations can then
be analyzed quantitatively through analytical and numerical computations.
Having a time-frequency analysis and a quantum optics 
perspective on the problem of relativistic detector response provides
a synthetic approach to the problem of moving detectors.
Besides, this time-frequency perspective sheds new lights on the 
interpretation of the measured signal and the problem of defining a 
notion of particles. Indeed, having two natural ways of defining 
particles, the standard many-body one and the operational one, 
demands to relate them and understand their interplay. Time-frequency
analysis offers a way to define relative stationary timescales from 
which notions of particles can be defined locally in spacetime and 
frequency domains.

This paper is structured as follows. In \cref{sec/context} we set up the
general framework of correlation functions and their time-frequency 
representation through the Wigner function. In \cref{sec/vacuum},
we analyze the response of a detector probing the vacuum for 
different non-stationary motions of the detector. We give a detailed
analysis of the adiabatic regime, its corrections and its breakdown.
\Cref{sec/excess} is dedicated to the study of the detector’s 
response in the presence of excitations in the uniformly accelerated
and realistic motions. In particular, how coherences in a superposition
transform can be analyzed straightforwardly. 
We conclude this paper in \cref{sec/bigpicture} by discussing how 
different notions of particles can be defined from the signal 
from a time-frequency analysis.

\section{First order correlation} 
\label{sec/context}

\subsection{Context and photo-detection}

Systems in quantum optics, condensed matter and high-energy 
physics are well described using the framework of quantum field 
theories. In this context, the experimentally relevant  quantities
are not the fields themselves but correlations functions constructed
from them. Some of them are known in quantum optics as coherence or 
Glauber functions. They naturally come by when analyzing the 
photo-detection response.

We are also here interested in the photo-detection response of 
a system moving arbitrarily in flat spacetime. It is designed to
detect a single excitation of a relativistic quantum field. We suppose 
that this device is moving in Minkowski spacetime with a given 
trajectory $\mathbf{x}(\tau)$ and is coupled linearly to a massless
scalar field $\phi(x)$. In the inertial laboratory reference frame the
Hamiltonian is given in the interaction picture by:
\begin{align}
	H_I(\tau) = d(\tau) \cdot \phi^{(I)}(t(\tau),\mathbf{x}(\tau))
	\,.
\end{align}
If we model the detector as a two level system of energy $\omega_{eg}$,
then $d(\tau)= -g \sigma_x(\tau)$  with $g$ the coupling constant.

We are now interested in the probability to measure the excited state after a time $\tau$.
Since the coupling is weak, we can use time-dependent perturbation theory,
expand the evolution operator at the first order and obtain the desired 
probability $p_{\omega_{eg}}(\tau)$:
\begin{align}
	p_{\omega_{eg}}(\tau)
	=
	\left(\frac{g}{\hbar}\right)^2
	\int_0^\tau
		\me^{\mi \omega_{eg} (\tau_1-\tau_2)}
		G(\tau_1,\tau_2)
		\,
	\md \tau_1 \md \tau_2
	\,.
\label{eq:photodetection-proba-2level}
\end{align}
The function $G(\tau_1,\tau_2)$ depends only on the state of the scalar 
field and is defined as a first order correlation function:
\begin{align}
\label{eq:photodet-correlation}
	G_\rho(\tau_2,\tau_1) = 
	\trace\big( \phi^{(I)}(t(\tau_1),\mathbf{x}(\tau_1))
	\phi^{(I)}(t(\tau_2),\mathbf{x}(\tau_2)) \rho \big)
	\,.
\end{align}
This correlation function contains all the contribution of the 
field to first order in the photo-detecting signal. 
There is however a major difference between this signal and 
the standard one found by Glauber in quantum optics. Indeed,
for photons, we have 
$
G^{\text{ ph}}_\rho(\tau_2,\tau_1) = \trace \big( E^-(\tau_2) E^+(\tau_1) \rho \big)
$
where $E^\pm$ are the positive and negative frequency 
parts of the electric field operator. In the relativistic regime,
the correlation function does not only depend on the product
$\phi^-\phi^+$ but on the full field as in \cref{eq:photodet-correlation} 
\cite{Svaiter-1992}. One reason behind this 
difference is fundamental and comes from the fact that the 
definition of positive and negative frequencies depends on 
the time coordinate. For a detector in a general trajectory 
or in the presence of a gravitational field, there is no 
global definition of time coordinate and so no general
decomposition of the field in momentum space. The 
notion of excitation and of vacuum become relative 
concepts. 

Equation \labelcref{eq:photodetection-proba-2level} 
can be generalized by introducing a generic linear response
function $\chi(\tau_2,\tau_1)$ of the detector and the 
resulting photo-detection signal is then obtained by 
\begin{align}
	p(\tau)
	=
	\int_\mathbb{R}
		\chi_\tau(\tau_2,\tau_1)
		G(\tau_1,\tau_2)
		\,
	\md \tau_1 \md \tau_2
	\,.
\end{align}
The function $\chi_\tau(\tau_2,\tau_1)$ characterizes the response of 
the detector and its form depends on the type of detector we use.
The photo-detection probability is then just the scalar product between
this response function and the first order correlation function.
 For a broadband device, the response will be local in time with
$\chi(\tau_2,\tau_1)=f(\tau) \delta(\tau_2-\tau_1)$ and $f(\tau)$ the 
switching function. On the contrary, for a narrow-band device like the 
two-level system, we measure the Fourier transform of the correlation 
function.

\subsection{Definitions}

\subsubsection{First-order and excess correlations}

The photo-detection problem shows that the quantity encoding the 
response of point-like detector at first order, is given by a 
first-order correlation function of the field defined as:
\begin{align}
G_\rho(\tau_2,\tau_1) 
	= \trace\big( \phi(\tau_1) \phi(\tau_2) \rho \big)
	= \langle \phi(\tau_1) \phi(\tau_2) \rangle_\rho
\end{align}
with the notation $\phi(\tau) = \phi^{(I)}(t(\tau),\mathbf{x}(\tau))$
for a given trajectory $x(\tau)$. Depending on the 
context, this function and all the higher-order ones that 
could be defined are called correlation functions or Wightman's 
functions. In quantum optics, the term coherence functions is used but 
involves correlation functions of the positive and negative frequency parts
of the field. We will stick to general quantum field theory denomination
of correlation functions. From now on, we use from now on a unit system
in which $\hbar = c = 1$ and the Minkowski metric signature
$(-,+,+,+)$.

The most important situation is when the vacuum state of the field is 
prepared. The correlation function can be computed exactly and is given 
by:
\begin{multline}
G_{\ket{0}}(\tau_2, \tau_1) = \\
	\frac{1}{4\pi^2}
	\frac{1}{-(t(\tau_1)-t(\tau_2)+\mi\epsilon)^2 + 
	(\mathbf{x(\tau_1)}-\mathbf{x(\tau_2)})^2}
\,.
\label{eq:correlation-inertial-vacuum}
\end{multline}
where, for the moment, we used the standard regularization of $\mi \epsilon$. 
The question of regularization will be discussed in more details 
in the next section.

Let's now add an extra-excitation in a normalized wave packet~$\Phi$:
\begin{align}
\phi[\Phi]\ket{0} 
	= \int_{\mathbb{R}^3} \Phi(t,\mathbf{x}) \phi^\dagger(t,\mathbf{x}) \ket{0} 
	\,\md^3 x
\,.
\end{align}
By Wick's theorem, and using the notation $\Phi^*\big(\mathbf{x}(\tau),t(\tau)\big) = 
\Phi(\tau)$, the first order correlation now reads:
\begin{align}
G_{\phi[\Phi]\ket{0}}(\tau_2, \tau_1)  
	&= G_{\ket{0}}(\tau_2, \tau_1) \nonumber \\
	&+ \Phi^*(\tau_1)\Phi(\tau_2) + \Phi(\tau_1)\Phi(\tau_2)
	+ \text{h.c.} 
\label{eq:correlation-1-particule}
\end{align}
This suggests to decompose the correlation function into two 
parts by the equation:
\begin{align}
G_\rho(\tau_2,\tau_1) 
	= G_{\ket{0}}(\tau_2, \tau_1) + \Delta G_\rho(\tau_2,\tau_1)
\,.
\label{eq:excess-correlation}
\end{align}
The interpretation is intuitively clear in the pure state case
described by \cref{eq:correlation-1-particule} since we can 
clearly think of excitations over the vacuum. For a 
general density matrix, the decomposition in 
\cref{eq:excess-correlation}
comes from the fact that a measurement must be understood as a
comparison between the state of the field and a reference state, which in
this case is the vacuum. This choice is also justified by the fact that
a reasonable physical state will have the same behavior as the vacuum
at high energy, both for absorption and emission processes. This turns
out to be important for the regularization aspects, as we will see.
However, such a decomposition might be more subtle when taking into
account general relativity and backreaction effects.

In the following, \cref{sec/vacuum} will focus on the vacuum 
contribution while the \cref{sec/excess} will be dedicated to the 
study of different kinds of excitation.

\subsubsection{On regularization}
\label{sec-reg}

In quantum field theory, the correlation functions are actually not 
proper functions but Lorentz-invariant distributions on spacetime
\cite{Wightman-2016}. The distribution character comes from the 
necessary divergences of the correlation functions which need
to be properly regularized.

The standard $-\mi\epsilon$ regularization procedure, that was used 
for instance in \cref{eq:correlation-inertial-vacuum}, corresponds to an
ultraviolet cut-off for the detector. However, in a general reference
frame, the frequency content is redistributed and some care must be
taken to ensure the proper regularization. A natural choice is to 
do a high-energy cut-off regularization similar to the $-\mi\epsilon$ 
regularization of the modes in the proper reference frame 
of the detector~\cite{Langlois-2006}.

This turns out to be equivalent to spatial regularizations, with
spatially-extended detectors
\cite{Schlicht-2004,Louko-2006,Satz-2007,Obadia-2007}, that were
introduced to solve the issues encountered with causality
leading to the impossibility to recover the Unruh effect 
with a causal detector~\cite{Schlicht-2004}, under the standard
regularization scheme.

All those regularization procedure are equivalent and lead to a 
well defined Lorentz invariant and causal correlation functions.
They amount to subtract the vacuum contribution found by an 
inertial detector~\cite{Louko-2006}. In the end, this strategy 
matches the one used in quantum optics and condensed
matter. The rationale behind it is physically intuitive, because the
correlator itself is not probed directly, but always compared to the
one of a reference state, as defined in \cref{eq:excess-correlation}.

\subsection{Representations of the first order correlation}

\subsubsection{Time and frequency representations}

The time representation $G_\rho(\tau_2,\tau_1)$ is the natural 
representation to look for dynamical information. The diagonal 
$G_\rho(\tau,\tau)$ corresponds to an energy density per unit time while 
the off-diagonal elements, which are complex numbers, give the coherences 
in time. However, this representation is not well suited to understand 
the kind of processes happening in the detection events since they are 
encoded in the $\tau_1-\tau_2$ dependence of the phase of 
$G_\rho(\tau_2,\tau_1)$. 

\begin{figure}
	\begin{center}
		\includegraphics{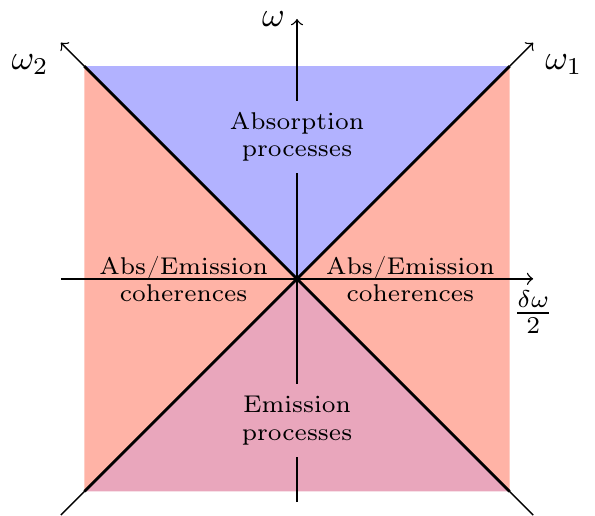}
	\end{center}
	\caption{%
		Decomposition of the Fourier plane into four quadrants: the
		upper quadrant corresponds to absorption processes, the lower
		quadrant to emission processes and the side quadrants to the
		coherences between emission and absorption processes.  While
		this particle-like interpretation makes sense for inertial
		observers, it does not necessarily hold as such for any
		trajectory.
	}
\label{fig/fourier-representation} 
\end{figure}

This is solved by going to the frequency domain. By computing a double 
Fourier transform, we can then define
\begin{align}
G_\rho(\omega_2,\omega_1) 
	&= \int_{\mathbb{R}^2} 
		G_\rho(\tau_2,\tau_1)  
		\, \me^{\mi (\omega_1 \tau_1 - \omega_2 \tau_2)}
	\, \md \tau_1 \md \tau_2 \nonumber \\
	&= \langle \phi^\dagger(\omega_1) \phi(\omega_2) \rangle_\rho
\end{align}
where the field $\phi(\omega)$ is defined with respect to an inertial mode 
decomposition as
\begin{align}
\label{eq/fourier-field}
\phi(\omega) = 
\int_{\mathbb{R}^3}
\left(
a_{\mathbf{k}}f_\mathbf{k}^*(\omega) + 
a^\dagger_{\mathbf{k}}f_\mathbf{k}(-\omega)
\right)
\;
\frac{\md^3\mathbf{k}}{2\omega_\mathbf{k} (2\pi)^3}
\end{align}
with 
\begin{align}
f_\mathbf{k}^*(\omega) = \int_\mathbb{R}
\me^{\mi k \cdot x(\tau)} \me^{\mi \omega \tau}
\, \md\tau
\,.
\end{align}
The Fourier plane $(\omega_1,\omega_2)$ is traditionally divided into
four quadrants as shown in \cref{fig/fourier-representation}.
The positive frequency quadrant, defined by $\omega_1>0$ and 
$\omega_2>0$, corresponds to the absorption processes while the negative
frequency quadrant corresponds to the emission processes. Finally, the 
two quadrants defined by $\omega_1\omega_2 <0$ correspond to the
coherences between emission and absorption processes. This 
interpretation follows the operational definition of particles and matches
the many-body one for inertial detectors. This equivalence does not hold 
for a general moving detector since nothing guaranties that the same 
notion of particles exists in all frames. Still we can expect that the
different notions of particles that we could be defined should match at
sufficiently high frequency (compared to acceleration or local
curvature). This is corroborated for a uniformly accelerated detector:
the inertial modes $u^{(i)}_\omega$ and  the accelerated modes 
$u^{(a)}_\omega$ are related by a Bogoliubov transformation
$u^{(a)}_\omega = (u^{(i)}_k 
- \me^{-\pi \omega/a} \bar{u}^{(i)}_k)
/\sqrt{1-\me^{-2\pi\omega/a}}$
from which we clearly see that for $\omega \gg a$, 
$u^{(a)}_\omega \approx u^{(i)}_k $. This remark suggest that
instead of trying to define a global notion of particle, we should 
maybe seek to define local notions of particles relative to the 
different scales of the problem: this hints toward a time-frequency
definition of particles, an idea that will be discussed in more 
details in \cref{sec/bigpicture}.

The diagonal $G_\rho(\omega,\omega)$ corresponds to the excitation 
occupation number per frequency. A convenient representation of the 
Fourier plane, also shown in \cref{fig/fourier-representation}, is given
by the variables $\delta\omega = \omega_1 - \omega_2 $ and 
$\omega = (\omega_1 + \omega_2)/2$ conjugated respectively to 
$\tau_1-\tau_2$ and $(\tau_1 + \tau_2)/2$ which, as we will see, are the natural 
variables for the time-frequency Wigner representation.

The frequency domain representation has complementary advantages
compared to the time representation. When analyzing the response
of a detector in a stationary trajectory, choosing one representation 
over the other is a matter of convenience. However, most physically 
realizable motions are not stationary and a time-frequency 
representation is called for. Such representations 
exist and have been analyzed in depth in signal processing research
\cite{Flandrin-1998}. The common one used in physics is the Wigner
representation which we will discuss in the context of relativistic 
field theory.

\subsubsection{Time-frequency representation}

The time and frequency representations have complementary properties:
while one clearly represents the time evolution, the other clearly shows
the type of processes taking place. While this is not a major issue for 
stationary signals, it becomes one for non-stationary signal like those
obtained by a detector moving in a general trajectory. Fortunately, 
it is possible to have the best of both worlds in one clear time-frequency 
representation. We propose to analyze the Wigner representation 
of the correlation function defined as:
\begin{align}
W_\rho(\tau, \omega) = \int_\mathbb{R}
G_\rho(\tau + \upsilon/2,\tau - \upsilon/2) 
\, \me^{\mi\omega\upsilon}
\,
\md\upsilon 
\,.
\end{align}
In the same way as \cref{eq:excess-correlation}, we can define 
an excess Wigner function $\Delta W_\rho$ with respect to the
vacuum defined as:
\begin{align}
W_\rho(\tau, \omega) 
	= W_{\ket{0}}(\tau, \omega) 
	+ \Delta W_\rho(\tau, \omega) 
\,.
\label{eq:excess-wigner}
\end{align}
The vacuum Wigner function must be regularized. As argued in 
\cref{sec-reg}, this is done by properly analyzing the response in 
the vacuum of an inertial detector and subtracting it. The Wigner function 
$ W_{\ket{0}}$ is again decomposed into two contributions 
\begin{align}
W_{\ket{0}}(\tau, \omega)
	= W^{\text{in}}_{\ket{0}}(\tau, \omega) 
	+\Delta_{\text{in}} W_{\ket{0}}(\tau, \omega)
\,.
\end{align}
The first one, $W^{\text{in}}_{\ket{0}}$, is the divergent inertial 
contribution which can be evaluated easily as 
$W^{\text{in}}_{\ket{0}}= \frac{|\omega|}{2\pi} \Theta(-\omega)$.
The second, $\Delta_{\text{in}} W_{\ket{0}}$, is the regular part that 
encodes the non-inertial contributions. It is a Fourier transform of
\cref{eq:correlation-inertial-vacuum} (without the $\mi\epsilon$
regularization) defined by
\begin{align}
\Delta_{\text{in}} W_{\ket{0}}(\tau, \omega) = 
	\frac{1}{4\pi^2} 
	\int_{\mathbbm{R}}
		\left(
			\frac{1}{(\Delta x (\tau,\upsilon))^2} - \frac{1}{-\tau^2} 
		\right)
		\me^{\mi\omega\upsilon}
	\,
	\md\upsilon 
\label{eq:vacuum-wigner-noninertial}
\end{align}
with $\Delta x (\tau,\upsilon) 
= x(\tau + \upsilon/2) - x(\tau - \upsilon/2)$.

The simplest situation is of course to consider an inertial detector in 
the vacuum. The response of the detector will then be given by
$W(\tau, \omega) = \frac{|\omega|}{2\pi} \Theta(-\omega)$.
This Wigner function is independent of $\tau$ which is 
a natural consequence of the stationary character of the
trajectory. Its form could have been anticipated by remembering
Fermi's Golden rule which states that the transition rate is given to 
first order by $2\pi d(\omega) f(\omega)$ with $d(\omega)$ the density
of states and $f(\omega)$ the distribution both in energy space. 
For our relativistic set up, the relativistic density of state is given
by
$\md^3 k / 2k^0 (2\pi)^3 = \omega \md \omega / 4\pi^2$ 
since in the massless case $\omega = |\mathbf{k}|$.

Non-trivial physics is unraveled for a detector in a uniformly 
accelerated motion. Indeed, consider the trajectory to be 
$x(\tau) = ( a^{-1}\sinh(a\tau), a^{-1}(\cosh(a\tau)- 1 ))$. Then the 
now well known thermal response is obtained:
\begin{align}
\Delta_{\text{in}} W_{\ket{0}}(\tau, \omega)
	= \frac{\omega}{2\pi} \frac{1}{\me^{2\pi\omega/a}-1}
\,.
\end{align}
While still stationary as expected, the Wigner function does 
not vanish for positive omega. This comes from the mixing of positive
and negative frequencies between the inertial and uniformly 
accelerated modes. The response of the detector is the same as a
thermal state with a temperature given by (using SI units)
\begin{align}
T = \frac{\hbar}{c k_B} \frac{a}{2\pi}
\,.
\label{eq/temperature}
\end{align}


The Wigner function possesses a nice set of properties. First, 
for a stationary signal like the previous examples, the Wigner 
function is time independent and positive. Moreover, the Wigner 
function possesses a frequency symmetry $W(\tau,\omega) =
W(\tau,-\omega)$ coming from the Hermitian property of the field.
 Second, its marginals give access to the probability distribution 
 of the conjugated variable. For instance, averaging over time 
 gives the spectral energy density distribution
\begin{align}
f(\omega) = \overline{W(t,\omega)}^t
\,.
\end{align}
In the $T$-periodic case, this average is taken over a time period,
implying
$f(\omega) = \frac{1}{T} \int_{-T/2}^{T/2} W(\tau,\omega) \md
\tau$.
Similarly, the integration over frequency gives the power $P(\tau)$,
which is finite only for the regularized Wigner function:
\begin{equation}
	P(\tau) = \int \Delta_{\text{in}} W_\rho(\tau, \omega)
	\, \frac{\md \omega}{2 \pi} = \Delta_{\text{in}} G_\rho(\tau,\tau)
\,.
\label{eq/def/power}
\end{equation}
This quantity has an interesting relation to the trajectory of the 
detector in the one dimensional case as we will see later and it
was proposed to use it as a general definition of temperature 
in curved spacetime \cite{Buchholz-2007,Buchholz-2012}.
Finally, the average over time and positive frequency gives
back the average energy measured by the detector
\begin{equation}
	\langle E \rangle_\rho = \int_{\mathbb{R}\times [0,+\infty[} 
	\Delta_{\text{in}} W_\rho(\tau, \omega)
	\, \md\tau \, \frac{\md \omega}{2 \pi}
\,,
\label{eq/def/averageenergy}
\end{equation}
an important property to keep in mind to normalize the states
we will consider.

\subsubsection{On causality}

Many different kinds of time-frequency representations exist and 
have been analyzed in the signal processing literature 
\cite{Flandrin-1998}. They can be classified according to a set of 
natural properties we could demand for a good representation of 
physical processes: unitarity (a measurement result translates
as a scalar product for representations), marginals corresponding to
spectral density and power spectrum, positivity (negativities
prevents probabilistic interpretations), linearity (a linear filter
translates as a linear filter for the representations),
causality and time-reversal symmetry. However, it happens that it is not
possible to construct a function satisfying all those requirements.
\Cref{table/pagevswigner} shows the properties of two important
time-frequency distributions.

Up to now, in the context of point-like detectors probing a relativistic
quantum field, only the Page distribution, which is a causal 
time-frequency distribution, has been studied
\cite{Schlicht-2004,Louko-2006,Satz-2007}.
Indeed, the main motivation was to understand if
the thermal behavior
would appear in a causal response which is not how the  standard
Unruh effect is derived. 

While this is more natural, the Page distribution is not convincingly
more physical than a non-causal one since we still integrate over the
whole past history of the motion. Indeed, a true physical response
is causal and happens during a finite duration. This is properly modeled
by considering a causal switching function $\chi_\tau(\tau_2,\tau_1)$
with finite support. By putting causality considerations in the
switching function, focusing on the Page distribution is not mandatory
anymore. It is even more interesting to consider the Wigner 
distribution, containing the same information as the Page one, since it
has a clearer interpretation. First, time-reversal symmetry in the
physical processes will be properly represented by the Wigner
distribution. Moreover, the Wigner function possesses the
linearity property which means that the Wigner transform of a 
linearly filtered signal is simply the scalar product between the Wigner
functions of the filter and the original signal. This is a clear advantage
over the other distribution for both signal processing
tasks and interferometric experiments. 
\begin{table}[h]
\caption{Comparison of the properties of the Page and Wigner
distributions.}
	\label{table/pagevswigner}
\centering
	\def\cmark{\ding{51}}
	\def\xmark{\ding{55}}
	\renewcommand*{\arraystretch}{1.1}
	\setlength{\tabcolsep}{4pt}
\begin{tabular}{@{}lcc@{}}
\toprule
	\multicolumn{1}{c}{\textbf{Properties}} & \textbf{Wigner} & \textbf{Page} \\ 
\midrule
	Unitarity & \cmark & \cmark  \\
	Positivity & \xmark & \xmark  \\
	Marginals & \cmark & \cmark \\
	Linearity & \cmark & \xmark \\
	Causality & \xmark & \cmark \\
	Time reversal & \cmark & \xmark \\
\bottomrule
\end{tabular}
\end{table}

\section{A detector in the vacuum} 
\label{sec/vacuum}

Let us now discuss the response of a detector probing the inertial
vacuum. The main purpose here is to understand the structure of the 
Wigner function of the vacuum for a generic trajectory. After setting up
the framework, we will first analyze the slow deviations from the 
uniformly accelerated case corresponding to the adiabatic approximation.
We will then discuss its breakdown by analyzing oscillatory motions in
the vacuum to finish with more physically realizable motions. 

\subsection{General 1+1D motion}
\label{sec/general-motion}

To simplify the theoretical analysis, we will consider here a  1+1D 
generic motion. The solution of the special relativistic equations of 
motion for a detector can be parametrised in a transparent way.
Starting from the normalization condition on the 4-velocity 
$-u_t^2 + u_x^2 = -1$, we have the natural parametrization:
\begin{equation}
	\mathbf{u}(\tau) = 
	\begin{pmatrix}
		\cosh A(\tau) \\[2pt]
		\sinh A(\tau)
	\end{pmatrix}
\,.
\end{equation}
By denoting $a(\tau)$ the oriented norm of the 4-acceleration 
$a^\mu(\tau)$(positive if the acceleration goes towards $x > 0$, 
negative otherwise) and re-injecting into the equation for the 
4-acceleration, we find that $a^2 = (\partial_\tau A)^2$ and, 
in the locally inertial frame at $\tau = 0$,  we have
\begin{equation}
	\mathbf{x}(\tau) =
	\begin{pmatrix}
		\int_0^{\tau} \cosh A(\tau') \, \md \tau'
		\\[1ex]
		\int_0^{\tau} \sinh A(\tau') \, \md \tau'
	\end{pmatrix}
\end{equation}
where $A(\tau) = \int_0^\tau a(\tau') \, \md \tau'$.
From this we can go one step further and express 
$\Delta x^2(\tau+\upsilon/2, \tau-\upsilon/2)$ in a suitable form
for analytical and numerical analysis. For that, we introduce the
quantity:
\begin{equation}
	A_\tau(\upsilon) = \int_\tau^{\tau+\upsilon} a(\tau') \, \md \tau'
	= A(\tau+\upsilon) - A(\tau)
\,.
\end{equation}
We then have:
\begin{multline}
	\label{eq/Dxsq/linacc/cosh}
	\Delta x^2(\tau+\upsilon/2, \tau-\upsilon/2)
	=\\
	- \int_{-\tau/2}^{\tau/2}
	\cosh(A_\tau(\tau_1) - A_\tau(\tau_2)) \, \md \tau_1 \md \tau_2
\,.
\end{multline}
Using the $\cosh$ definition, we can see that this double integral can be
re-expressed as a product of two simple ones:
\begin{equation}
	\Delta x^2(\tau+\upsilon/2, \tau-\upsilon/2) =	
	- f_+(\tau,\upsilon) f_-(\tau, \upsilon)
\end{equation}
where $f_\pm(\tau, \upsilon) = \int_{-\upsilon/2}^{\upsilon/2}
\exp(\pm A_\tau(\upsilon')) \, \md \upsilon'$. We can also re-express the 
trajectory in terms of $f_\pm$. A description in the locally-inertial 
frame at time $\tau$ would simply be:
\begin{equation}
	\Delta \mathbf{x}_\tau(\upsilon)
	=
	\frac{1}{2}
	\begin{pmatrix}
		f_+(\tau, \upsilon) + f_-(\tau, \upsilon)\\[2pt]
		f_+(\tau,\upsilon) - f_-(\tau,\upsilon)
	\end{pmatrix}
\,.
\end{equation}
This expression in terms of $f_\pm$ possesses a few advantages. It is
centered around $\tau$, which allows to perform expansion for small values
of $\upsilon$. Conversely, it allows precise numerical evaluation around
small $\upsilon$ values, which is of prime importance in the regularization
scheme we have chosen. 

The Wigner function can then be computed using
\cref{eq:vacuum-wigner-noninertial}. An interesting property can already
be obtained for the power $P(\tau)$. Indeed, by computing
the two sides of \cref{eq/def/power}, we have:
\begin{equation}
	P(\tau) = \frac{1}{4\pi^2} \frac{a_\tau^2}{12}
\,.
\label{eq/power}
\end{equation}
The rationale behind defining local temperature in a general spacetime 
\cite{Buchholz-2007,Buchholz-2012} comes from this relation and the fact
that the acceleration for a  uniformly accelerated detector is
proportional to the temperature (\cref{eq/temperature}), a property 
that remains true for an adiabatic motion as we will now see.

\subsection{Adiabatic regime and its breakdown}
\label{sec/adiabatic}
\subsubsection{Adiabatic regime}

When acceleration changes slowly, we expect the Wigner function to be 
close to the uniformly accelerated case: this is called the adiabatic
regime \cite{Obadia-2007,Padmanabhan-2010,Visser-2011,Barbado-2012}. 
More precisely, we expect that the main contribution to the 
Wigner function to be similar to a thermal response with a time 
dependent temperature $T(\tau)$ proportional to the instantaneous 
acceleration $a(\tau)$. 

For the purpose of this discussion, we write explicitly the functional
dependence on the acceleration of the Wigner function as 
$W[a(\tau)](\tau,\omega)$. Given a time $\tau$, we denote
the uniformly accelerated trajectory having the acceleration 
$a(\tau)$ by $a_\tau$. Doing an expansion around this trajectory,
we obtain 
\begin{multline}
W[a(\tau)] = W[a_\tau] 
+ 
\int_{\mathbb{R}} 
\frac{\delta W}{\delta a(\upsilon)} [a_\tau] \, \delta a(\upsilon)
\, \md \upsilon 
\\
+
\frac{1}{2}\int_{\mathbb{R}^2}  
\frac{\delta^2 W}{\delta a(\upsilon_1) \delta a(\upsilon_2)}
[a_\tau] \, \delta a(\upsilon_1) \, \delta a(\upsilon_2)
\, \md \upsilon_1 \md \upsilon_2
\,.
\label{eq/functional-expansion}
\end{multline}
The first term corresponds to the adiabatic response of the 
detector: the Wigner function is the thermal distribution with 
a time-dependent temperature proportional to the instantaneous 
acceleration $a_\tau$. The other terms are corrections to this 
dominant term.

This development is meaningful when the variations of the acceleration
$\delta a_\tau(\upsilon)$ around a given time $\tau$ are small compared 
to the acceleration $a_\tau$ over a timescale 
$\tau_{\text{s}}\gg a_\tau^{-1}$:
\begin{align}
\delta a_\tau(\upsilon) \ll a_\tau
\text{ with }
\upsilon \le \tau_{\text{s}}
\,.
\label{eq/criterion}
\end{align}
The timescale $\tau_s$, that we could call adiabatic or stationary time,
is of prime important since it gives us the interval of time around $\tau$
over which we can consider the motion uniformly accelerated.
Moreover, the variation $\delta a_\tau(\upsilon)$ can itself be seen as a 
function of the derivative $(\dot{a},\ddot{a},\dots)$. In good 
regimes, it is legitimate to do an expansion in those derivatives and 
obtain the reduced and more familiar criterion $\dot{a}/a^2 \ll 1 $.

Thus both the amplitude and the frequency of the perturbation
play a role in defining the adiabatic regime and deviations form 
it. To properly understand the different regimes of the response, 
we consider an oscillatory acceleration of the form
${a(\tau) = a_0 + a_1 \sin(2 \pi f \tau)}$ with $a_0$ a constant 
acceleration and $(a_1,f)$ the amplitude and frequency of the 
oscillatory drive \cite{Doukas-2013}. 
The functional expansion of \cref{eq/functional-expansion} can then 
qualitatively be seen as an expansion in $a_1/a_0$ while the 
derivative expansion of $\delta a$ is an expansion in $(2\pi f)/a_0$. 
The different regimes can then be classified as follows:
\begin{itemize}
\item The adiabatic regime is valid when the perturbation is 
small such that $a_1 \ll a_0$ and $2 \pi f \ll a_0$. The thermal response
follows the acceleration as in $W[a_\tau]$ and is corrected by small
terms in the derivatives of the acceleration.
\item The adiabatic regime \emph{per se} breaks down when one of the 
two conditions above is not fulfilled and will be analyzed in the next
section. In the regime $a_1 \ll a_0$ and 
$f \gtrsim a_0$, the functional expansion still works but the terms rearrange
themselves such that a thermal response is still present at the average 
acceleration $\bar{a}=a_0$ plus corrections of order $1/f$.
\item Finally, in the regime $a_1 \gtrsim a_0$, all the 
expansions break down and the structure of the Wigner function 
has to be analyzed differently.
\end{itemize}

We concentrate first on the pure adiabatic regime where we have 
$a_1 \ll a_0$ and $f \ll a_0$. In this regime, the intuition of a thermal 
response following the evolution of the acceleration works.
Furthermore, the overall order of magnitude of a correction to 
$W[a_\tau]$ coming from the functional and derivative expansions is 
given by powers of the from $(2\pi f/a_0)^p \cdot (a_1/a_0)^q$. 
\Cref{table/order-magnitude} sums this up from the first few 
corrections.
\begin{table}[h]
\centering
\caption{\label{table/order-magnitude} Orders of magnitude (in units of 
$a_0$) of the corrections in the functional and derivative expansion.}
	\renewcommand*{\arraystretch}{1.3}
	\setlength{\tabcolsep}{3pt}
	\begin{tabular}{@{}rl@{\hskip 12pt} rl@{\hskip 12pt} rl@{}}\toprule
	\multicolumn{2}{c}{$O(\delta a)$}
	&
	\multicolumn{2}{c}{$O(\delta a^2)$}
	&
	\multicolumn{2}{c}{$O(\delta a^3)$}  \\ \midrule
	$(2\pi f)^2 a_1$ &  $[\ddot{a}]$
	& 
	$(2\pi f)^2 a_1^2$ & $[\dot{a}^2]$  
	&
	\multicolumn{2}{c}{--} \\
	\multicolumn{2}{c}{--}
	&
	$(2\pi f)^3 a_1^2$ & $[\dot{a}\ddot{a}]$
	& $(2\pi f)^3 a_1^3$ & $[\dot{a}^3]$ \\
	$(2\pi f)^4 a_1$ & $[a^{(4)}]$
	&
	$(2\pi f)^4 a_1^2$ & $[\ddot{a}^2]$
	& $(2\pi f)^4 a_1^3$ & $[\dot{a}^2\ddot{a}]$ \\
\bottomrule
\end{tabular}
\end{table}

\begin{figure}
	\begin{center}
		\includegraphics{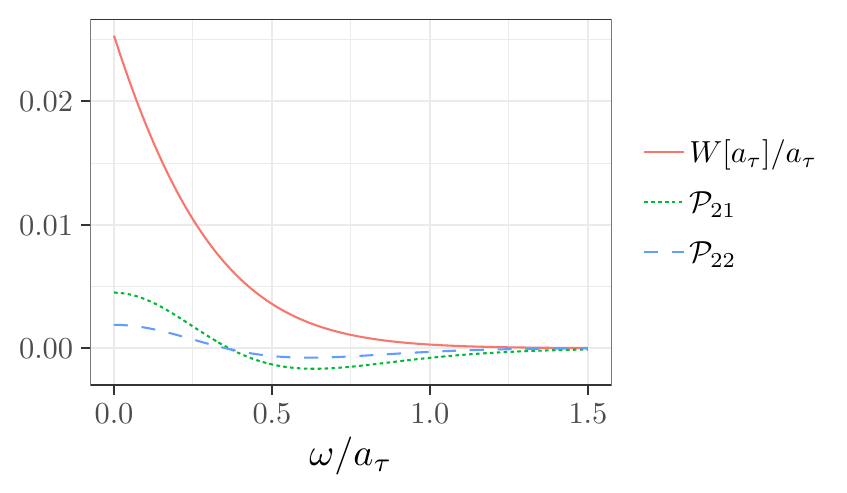}
	\end{center}
	\caption{Representation of the universal functions of the thermal
	distribution coming as corrections to the pure thermal response 
	$W[a_\tau]$ in the derivative expansion. Their form is 
	independent of the trajectory.
	\label{fig/polynomes}}
\end{figure}

\begin{figure}
	\begin{center}
		\includegraphics{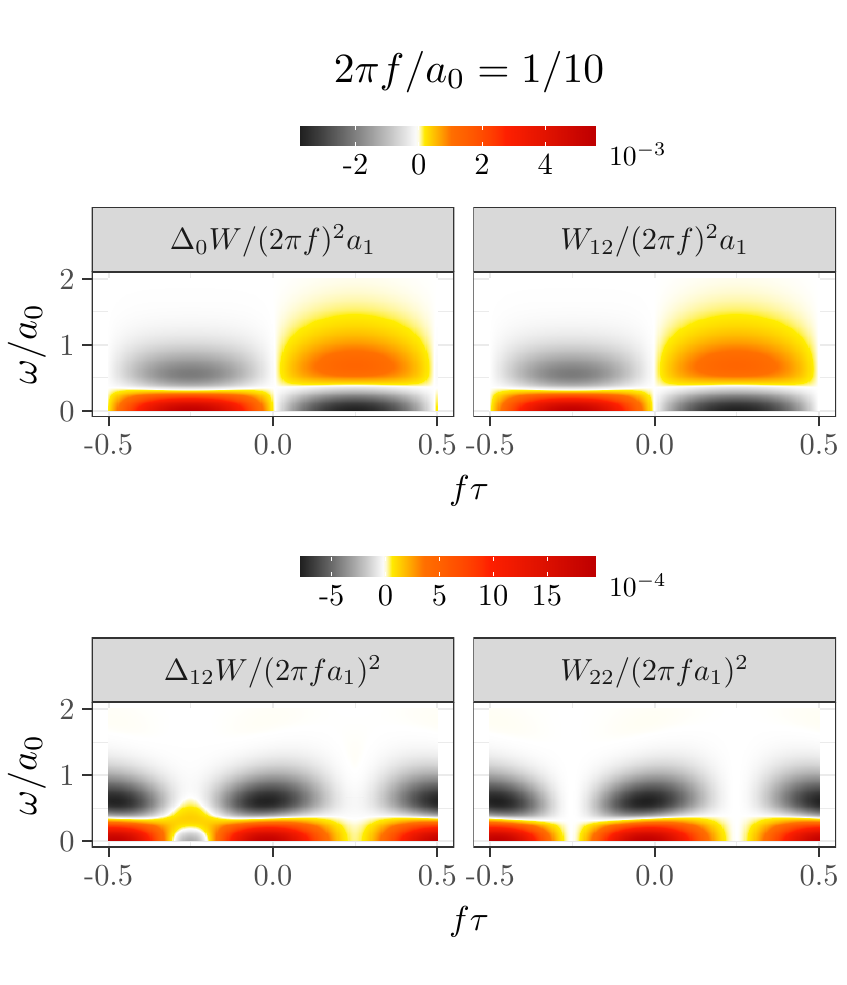}
	\end{center}
	\caption{Comparison between a correction to $W[a_\tau]$ at
	a given order $W_{ij}$ and the exact one at the same order
	${\Delta_{ij}W = W-\sum_{(k,l)<(i,j)} W_{kl}}$. In the adiabatic
	regime, at low frequency $f$, the derivative expansion is meaningful, 
	each corrections being a order of magnitude lower than the previous 
	one.
	\label{fig/sine-adiab}}
\end{figure}

Thanks to the symmetry of the Wigner function, the first correction
in \cref{eq/functional-expansion} only has even derivatives in $a$ in 
the derivative expansion. This means in particular that there is no 
$\dot{a}$ corrections to the thermal behavior. The first two corrections
to the Wigner function have the following form:
\begin{equation}
\begin{split}
W[a(\tau)] = W[a_\tau] 
&+\frac{\ddot{a}}{a^2}\mathcal{P}_{12}[g](2\pi\omega/a_\tau)
\\
&+\frac{\dot{a}^2}{a^3}\mathcal{P}_{22}[g](2\pi\omega/a_\tau)
\end{split}
\end{equation}
where ${g(x) =x/(\me^x-1)}$ is the thermal distribution and the 
$\mathcal{P}_{ij}\in\mathbb{R}[Y,X]$
are polynomials of two variables such that the action on $f$ is a 
derivative operation
$\mathcal{P}_{ij}[g] \equiv \mathcal{P}_{ij}(x,\partial_x)[g(x)]$.
Technical details about this derivation are given in 
\cref{sec/adiab-annex}.
\Cref{fig/polynomes} represents the two functions 
$\mathcal{P}_{12}[g](x)$ and $\mathcal{P}_{22}[g](x)$ which 
are universal in the sense that they do not depend on the 
trajectory of the detector while \cref{fig/sine-adiab} compares
each correction to the exact expression evaluated numerically
at a given order. This shows that the corrections to the 
adibatic thermal response are orders of magnitude less than
$W[a_\tau]$, thus justifying that the regime $a_1 \ll a_0$ 
and $f \ll a_0$ corresponds indeed to an adiabatic regime 
where the thermal response follows the evolution of the
acceleration.

\subsubsection{Breakdown of the adiabatic regime}
\label{sec/breakdown}

When the perturbation is too important, meaning that the conditions
$a_1 \ll a_0$ and $f \ll a_0$ are not both fulfilled, the adiabatic
response is not valid anymore. The simplest deviation we can first
consider is $f\gtrsim a_0$. Intuitively, we expect that, since the
frequency is too high, the thermal response cannot build up fast enough
and follow the variations of the acceleration. Only an average thermal
response at the acceleration $\bar{a}$ should build up while traces
of the oscillations should appear at higher frequencies in the 
time-frequency plane. This intuition can be explicitly checked by 
computing exactly the full first correction in \cref{eq/functional-expansion}
for the trajectory ${a(\tau) = a_0 + a_1 \sin(2 \pi f \tau)}$ denoted
$\Delta_0 W$. It actually contains all the derivative corrections
$a^{(n)}$ of order $n$ (first column of \cref{table/order-magnitude}).
Its explicit form is given by
\begin{multline}
W_1
=
\frac{a_1\sin(2\pi f \tau)}{4\pi^2}
\Big[
\frac{1}{1+(2\pi f /a_\tau)^2}\frac{g_+ + g_-}{2} - \\
\frac{\omega/2\pi f}{1+(2\pi f /a_\tau)^2}(g_+ - g_-) +
\frac{2\pi}{a_\tau}\omega \dot{g}_0-g_0
\Big]
\label{eq/firstordercorr}
\end{multline}
where we reused $g(x)$ the thermal distribution
and its values $g_\pm = g\left( 2\pi / a (\omega \pm \pi f) \right)$ and
$g_0 = g\left( 2\pi\omega / a \right) $.
It can be explicitly checked that in the limit $f \ll a_0$, we 
recover the $\ddot{a}$ correction to the adiabatic behavior.

From \cref{eq/firstordercorr}, we can now understand the high
frequency regime $f \gg a$. In the region $\omega \lesssim a_0$, we 
have 
\begin{align}
W_1 \approx  
\frac{a_1\sin(2\pi f \tau)}{4\pi^2}
\left[ 
\frac{2\pi}{a_\tau}\omega \dot{g}_0-g_0
\right]
\,.
\end{align}
This expression has a nice interpretation: by considering a uniformly 
accelerated trajectory $a_0$ perturbed by a constant small term
$a_1$, we have $W[a_0+a_1] = W[a_0] - \left[ 
\frac{2\pi}{a_\tau}\omega \dot{g}_0-g_0
\right]/4\pi^2$. Thus, we conclude that in the region $\omega \lesssim a_0$
in the high-frequency regime the full Wigner function has the simple 
expression 
\begin{align}
W[a(\tau)] = W[\bar{a}]
\,.
\end{align}
This matches the intuitive idea that the frequency of the perturbation 
is too high for a thermal behavior following the drive to build up.
In fact, by a proper expansion of \cref{eq/firstordercorr} in $2 \pi f$
(done in \cref{sec/adiab-annex}), we can see that there are corrections 
of order $1/2 \pi f$ to the average thermal response in the frequency 
band $\omega \in [0, \pi f[$.

\begin{figure*}
	\begin{center}
		\includegraphics{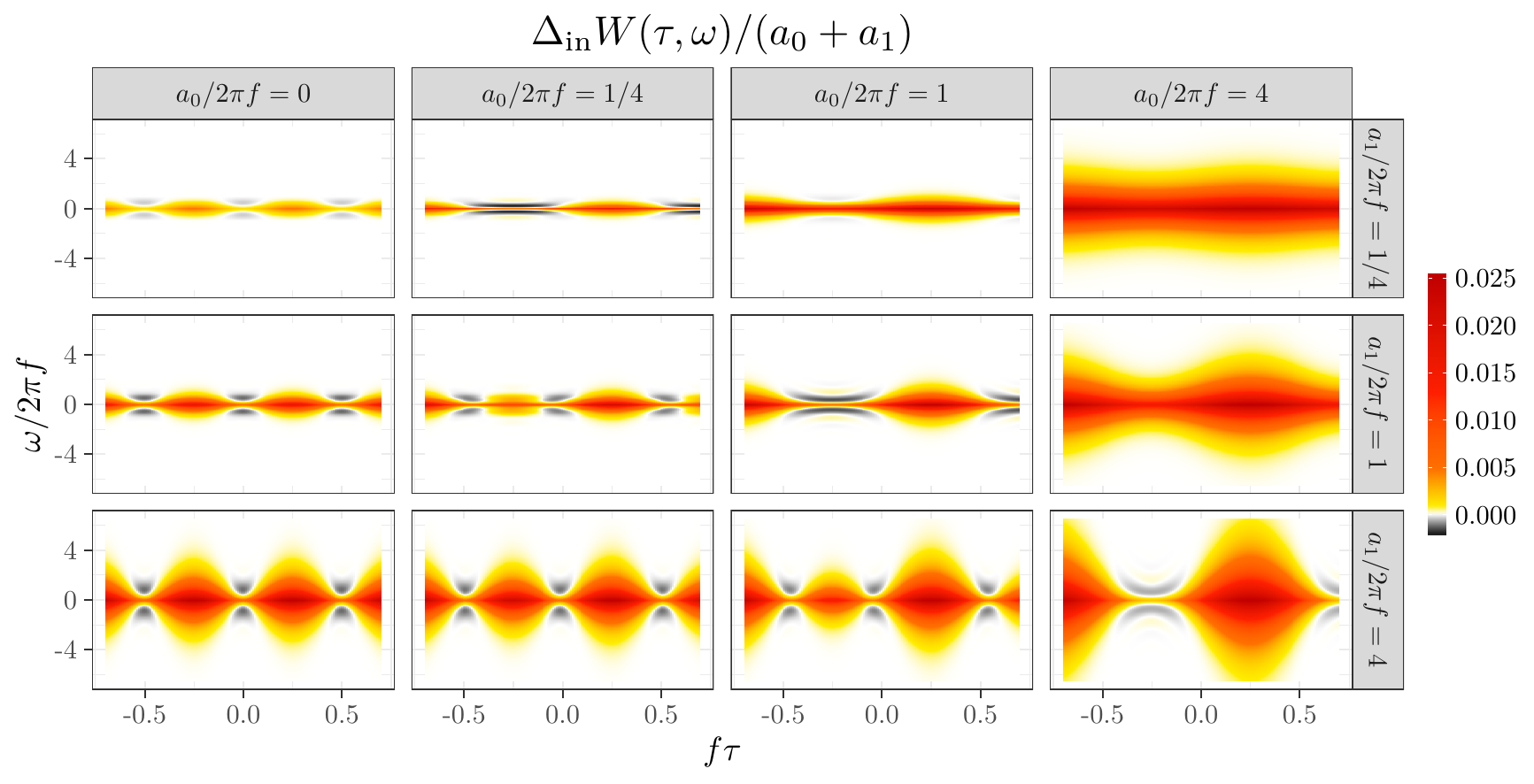}
	\end{center}
	\caption{Wigner function representation of an oscillatory acceleration 
	$a(\tau) = a_0 + a_1 \sin(2 \pi f \tau)$ in different regimes controlled
	by the expansion parameters $(a_0/2\pi f)$ and $(a_1/a_0)$. In the 
	regime of small frequency $f$ and amplitude $a_1$ compared to $a_0$,
	the adiabatic response works globally. Outside this regime, the adiabatic
	expansion breaks down, which is witnessed by the appearance of inner
	oscillations, but can still be meaningful locally.
	\label{fig/sinus}}
\end{figure*}

Finally, the expansion \labelcref{eq/functional-expansion} breaks down
completely when the criterion \labelcref{eq/criterion} is not satisfied.
In the oscillatory example $a(\tau) = a_0 + a_1 \sin(2 \pi f \tau)$ , 
this qualitatively means that $a_1 \sim a_0$. 
In fact, this characterization is too brutal
and global compared to the more local one from \cref{eq/criterion}: this 
means that globally the functional expansion cannot be performed but
it can remain meaningful in some time intervals.

\Cref{fig/sinus} represents the Wigner function of the oscillatory 
acceleration for different parameters $(a_0/2\pi f,a_1/2\pi f)$. The
global or local validity of the adiabatic expansion is witnessed by the 
appearance of inner oscillations in the Wigner function. In the 
regimes $(4,1/4)$ and $(4,1)$ for instance, the adiabatic expansion is
globally valid. This is not anymore the case for the other regimes where
$a_1 \sim a_0$ and where the signal basically goes (close) to zero 
at some moments in time. Still, the expansion remains meaningful 
locally half a period later. This can be made more quantitative 
by explicitly analyzing the criterion \labelcref{eq/criterion}. As an 
example, consider the situation where $a_1 = a_0$. The criterion 
is then equivalent to $\cos(2\pi f\tau+\pi f \upsilon) \sin(\pi f \upsilon) \ll 
\cos^2(\pi/4-\pi f t)$. Clearly, when $f\tau = -1/4$, the 
criterion cannot be satisfied and the adiabatic expansion breaks down
while it is valid around $f\tau=1/4$ (see \cref{fig/sinus}).
How can this be interpreted will be discussed in more details in 
\cref{sec/bigpicture}.

\subsubsection{A more physical trajectory}

The previous analyses, while important in their own regard to 
understand how the response changes in non stationary situations,
are still based on non physical trajectories since they require an
infinite amount of energy to be sustained.  The question then
remains on understanding the form of the Wigner function for 
physical trajectories \cite{Higuchi-1993,Padmanabhan-1996,Obadia-2007}.

\begin{figure}
\begin{center}
	\includegraphics{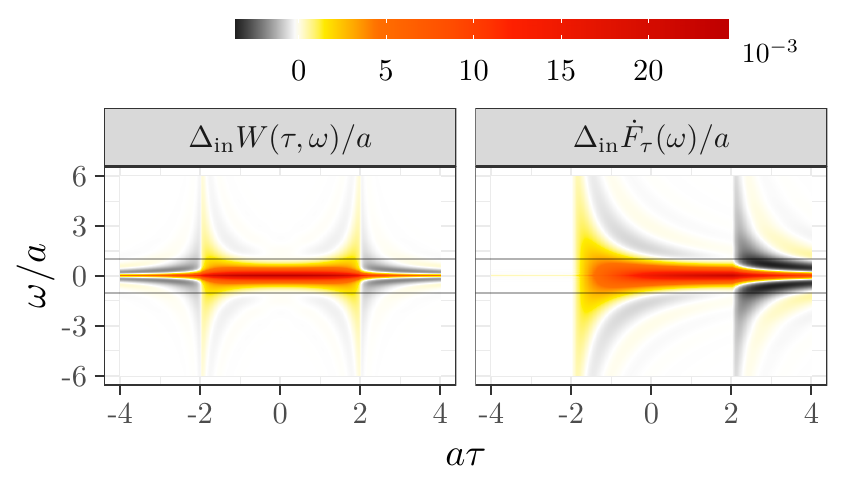}
\end{center}
\caption{Wigner (left) and Page (right) distributions of the vacuum for
a finite duration uniform 
acceleration between two inertial phases. After a transition time of the 
order of $a^{-1}$, the thermal behavior at temperature $a/2\pi$ settles 
down. The decreasing oscillating high-frequency parts are solely controlled
by the discontinuity of the acceleration.
}
\label{Wigner-accelerated-stopped} 
\end{figure}

\Cref{Wigner-accelerated-stopped} represents the Wigner function
(left panel) of a trajectory uniformly accelerated for a finite duration
$a\tau =4$. To make contact with the literature, it also shows 
(right panel) the Page distribution for the same trajectory. 
Besides the obvious causal response, we can see 
that a thermal response is building up over a timescale of a few $a$.
It is to be noted that the Page distribution, like the Wigner function, 
is not always positive in the time-frequency plane.

Concerning the Wigner function, its general features can be well understood.
First, we see that a thermal response at temperature $a/2\pi$ appears
over a timescale of the order of $a$. Second, the high-frequency 
structure around the beginning of the accelerated phase depends solely
on the discontinuity in the acceleration. In our case of interest,
we expect the second and higher derivatives of the first order 
correlation to be discontinuous. To analyze their effects on the 
Wigner representation, it is useful to use the following decomposition
$G(\tau + \upsilon/2, \tau - \upsilon/2)	= f_\tau(\upsilon) + 
g_\tau(\upsilon)$ where $f$ contains the lower order discontinuity
contribution and $g$ the higher order ones. The detailed form
of those functions are irrelevant for the high-frequency 
behavior and can be chosen for computational convenience: the only
constraints are that should capture the form of the  discontinuities
(see \cref{sec/discontinuity} for details on this strategy). 
In the end, we obtain the high 
frequency behavior of the Wigner function around the times
$\tau_d$ of brutal discontinuous changes of the acceleration as:
\begin{subequations}
\begin{align}
	\Delta W(\tau \ge \tau_d, \omega) &\simeq
	- \frac{1}{4\pi^2} \frac{a}{8 \sinh^2 a(\tau-\tau_d)} 
	\frac{\sin 2 \omega (\tau-\tau_d)}{(\omega/a)^3}
	\\
	\Delta W(\tau \le \tau_d, \omega) &\simeq
	- \frac{1}{4\pi^2} \frac{a}{16 a^3 (\tau-\tau_d)^3} 
	\frac{\cos 2 \omega (\tau-\tau_d)}{(\omega/a)^4}
\end{align}
\end{subequations}

\section{Excess correlation for different trajectories} 
\label{sec/excess}

Up to now, we have only been interested in the first order correlation of
the vacuum. We now move to the subject of the excitations above the
vacuum and how they are perceived by a moving detector 
\cite{Padmanabhan-2015}. From \cref{eq:correlation-1-particule}, the
excess correlation coming from a one-particle excitation in a 
wavefunction $\Phi(t,\mathbf{x})$ is given by
\begin{align}
\Delta G_\rho(\tau_2,\tau_1) = 
	\Phi^*(\tau_1)\Phi(\tau_2) + \Phi(\tau_1)\Phi(\tau_2) 
	+ \text{h.c.}
\,.
\end{align}
The nice feature of this correlation function is that its form is 
independent of the trajectory of the detector which is a direct
consequence of the covariance properties of the quantum field
correlation functions. 

The states mostly considered in a quantum optics setting are Fock 
states and coherent states. The main difference between the two
in the first order correlation is the absence (resp\@. presence) of 
the interference terms $\Phi(\tau_1)\Phi(\tau_2)+ \text{h.c.}$
for Fock states (resp\@. coherent states).

Finally, those states can be prepared in different wavepackets like a
monochromatic one or a Gaussian one. In what follows, we will mainly 
focus on Gaussian wavepackets of coherent and Fock states, which 
are in the end the most intuitive ones. We leave the mathematical 
analysis of the monochromatic case for \cref{sec/fock-coherent}.

\subsection{Gaussian wavepacket}

Let us consider, for now in 3+1D, that the inertial observer 
prepares the field in a Gaussian coherent state. It is defined in the 
following way:
\begin{align}
\ket{\alpha} = \bigotimes_\mathbf{p} \ket{\alpha_\mathbf{p}}
= \me^{\int_{\mathbb{R}^3}
	\big( \alpha_\mathbf{p} a^\dagger_\mathbf{p} 
	- \alpha^*_\mathbf{p} a_\mathbf{p}\big) \, \md^3 p }
\ket{0}
\,.
\label{eq/multimode-coherentstate}
\end{align}
The exponential operator is called the displacement operator $D(\alpha)$.
This state is normalized and satisfies the fundamental relations of coherent 
states\footnote{This is obtained from the BCH formula and the covariant 
commutation relations $[a_\mathbf{p},a^\dagger_{\mathbf{p}'}] = 
(2\pi)^3 2\omega_\mathbf{p} \, \delta(\mathbf{p}-\mathbf{p}')$.}
\begin{subequations}
\begin{align}
D(-\alpha) a_\mathbf{p} D(\alpha) &= a_\mathbf{p} + 
	(2\pi)^3 2\omega_\mathbf{p} \, \alpha_\mathbf{p} 
\,,\\
a_\mathbf{p}\ket{\alpha} &= 
	(2\pi)^3 2 \omega_\mathbf{p} \, \alpha_\mathbf{p} \ket{\alpha}
\,.
\end{align}
\end{subequations}
We can also think of this state in a spatial way by looking at its 
action on a field operator. Indeed,
\begin{align}
D(-\alpha) \phi^+(x,t) D(\alpha) = 
\phi^+(x,t) +
\Phi_\alpha(x,t)
\,.
\end{align}
where $\phi^+$ is the negative frequency part of the field.
Thus we see that the state \eqref{eq/multimode-coherentstate} 
is a coherent state in position with a parameter given by 
\begin{align}
\Phi_\alpha(x,t) = 
	\int_{\mathbb{R}^3}
		\alpha_\mathbf{p}
		\me^{-\mi (w_\mathbf{p} t - 
		\mathbf{p}.\mathbf{x})}
	\, \md^3 p
\,.
\end{align}
We see that at $t=0$ it is just the Fourier transform 
of the coherent state parameter in momentum space.
From the factorized nature of this state, the first order
correlation function can be decomposed into clear different
contributions
\begin{align}
\Delta G_{\ket{\alpha}}(\tau_2, \tau_1) &= 
\Phi_\alpha(\tau_1)\Phi_\alpha(\tau_2) +
\Phi_\alpha(\tau_1)\Phi_\alpha^*(\tau_2) \nonumber \\
&+ 
\text{h.c.} 
\,.
\end{align}

We now specify the function $\alpha_\mathbf{p}$ and choose
it so that the problem reduces effectively to a 1+1D problem
for computational simplicity. We then consider a Gaussian 
centered at a given momentum $p_0$, with a width given by 
$\sigma_p$. We then have
\begin{align}
\alpha_p = \sqrt{\frac{p_0}{2\pi}}\frac{1}{\left( 2\pi \sigma_p^2 \right)^{1/4}}
\me^{-\frac{(p- p_0)^2}{4\sigma_p^2}}
\me^{-\mi p.x_0}
\,.
\end{align}
In position it is a Gaussian centered at the position $x_0$.
We also make the following assumption that $p_0 \gg \sigma_p$ so 
that we can make consider that $\omega_p = |p| =p$ in our computations.
This leads to
\begin{align}
	\Phi_\alpha(\tau) = 
	\sqrt{\frac{p_0}{\left( 2\pi \sigma_x^2 \right)^{1/2}}}
	\me^{-\left[ (t_\tau - x_\tau)  + x_0 \right]^2/4\sigma_x^2}
	\me^{-\mi p_0\left[ (t_\tau - x_\tau) + x_0 \right]}
\,.
\label{eq/Gaussianwavepacket}
\end{align}
The introduction of the normalization $\sqrt{p_0/2\pi}$ comes 
from dimensional considerations since we require the average 
energy to equal the average value of the Wigner function over
time and frequency (see \cref{eq/def/averageenergy}).
We also introduced the position width $\sigma_x$ satisfying the 
relation $\sigma_x \sigma_p = 1/2$. 

As an example, consider an inertial trajectory with a velocity $v$, 
the world-line is parametrized as $(\gamma \tau, \gamma v \tau)$.
Its Wigner function is:
\begin{align}
	&W^{(v)}(\tau,\omega) = 2 p_0
	\Big[
		\me^{-\frac{(\omega- D_v p_0)^2}{2 (D_v\sigma_p)^2}} +
		\me^{-\frac{(\omega+D_v p_0)^2}{2 (D_v\sigma_p)^2}} + 
		 \nonumber \\
		&2\cos\left( 2p_0(D_v\tau + x_0) \right)
		\me^{-\frac{\omega^2}{2 (D_v\sigma_p)^2}}
	\Big]
	\me^{-\frac{(D_v\tau + x_0)^2}{2 \sigma_x^2}}
\,.
\label{eq/gaussian-inertial-velocity}
\end{align}
The computation is straightforward  and the Wigner function is composed
of two symmetric Gaussian spots centered around the Doppler shifted 
frequency $ D_v p_0$ with their interference pattern. 

Gaussian spots are in fact the basic ``atoms'' of the Wigner function
and allow to understand the geometry behind this representation
\cite{Flandrin-1998}. The basic interpretative element that we need 
and that we see in \cref{eq/gaussian-inertial-velocity} is that the 
interference term of two Gaussian atoms is also a Gaussian spot 
located at the mid-point joining the center of the two atoms 
(here $\omega = 0$) and that the interference pattern oscillates 
in the orthogonal direction.

This discussion would be completely similar if, instead of Gaussian 
coherent states, we consider a Gaussian superposition of a Fock 
state of $n$ photons. The excess correlation is even simpler since 
the interference terms vanish: 
${\Delta G_{\ket{n_\alpha}}= n\Phi^*_{n_\alpha}\Phi_{n_\alpha}
+\text{h.c.}}$.

While those Wigner functions could have been guessed intuitively
for an inertial response, it is a non-trivial task to analyze the 
response to a Gaussian excitation from a moving detector 
for different accelerated trajectories.

\subsection{Accelerated Wigner function}

\subsubsection{Uniformly accelerated case}

\begin{figure*}
	\begin{center}
		\includegraphics{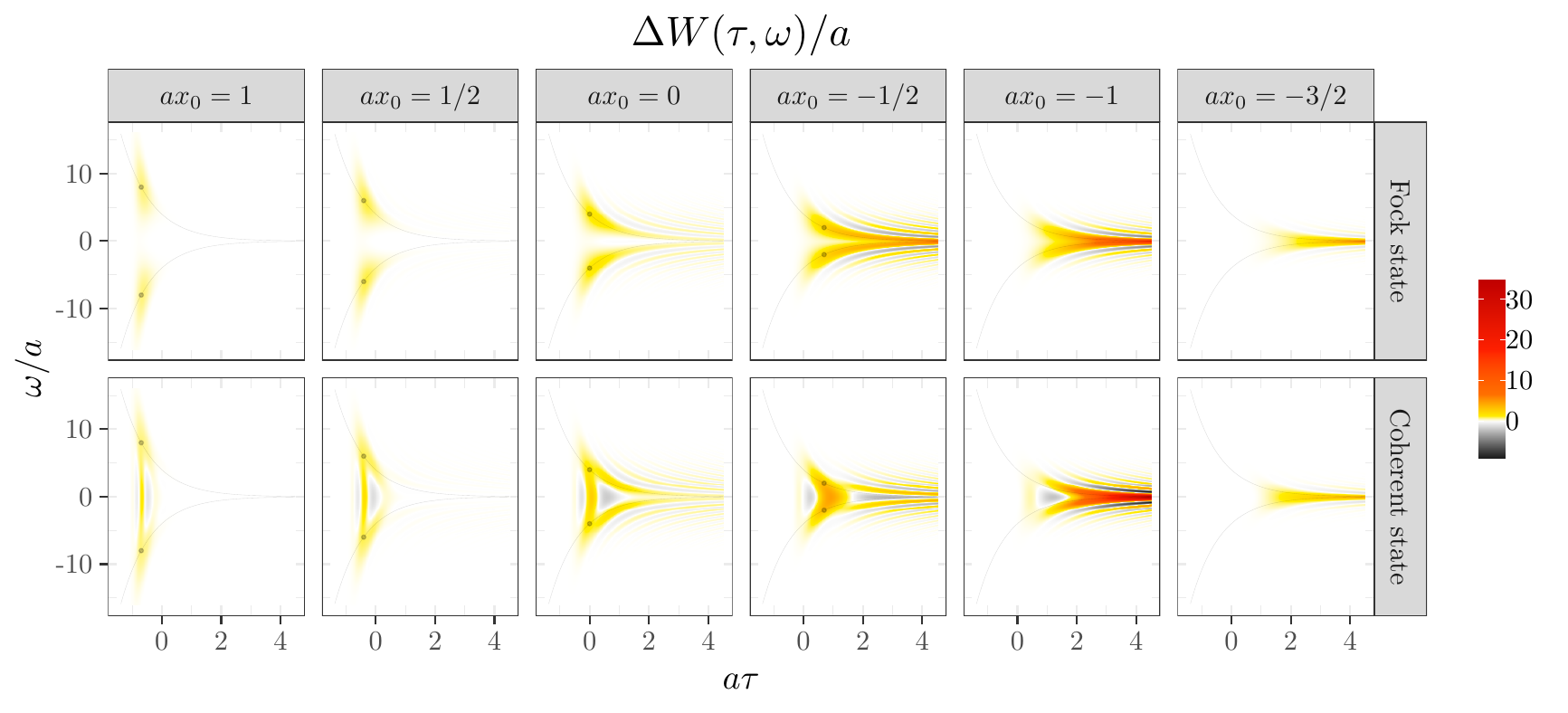}
	\end{center}
\caption{Evolution of the Wigner function of Gaussian Fock state (top row)
and coherent state (bottom row) emitted at different positions $x_0$, with
a frequency $p_0/a = 4$ and a width $a \sigma_x = 1/2$ in the inertial frame.
The signal is centered around a spot at $(\tau_r,\omega_r)$ given 
by \cref{eq/receptiontime-uni-acc,eq/receptionfreq-uni-acc} which 
are respectively the special relativistic reception time and frequency
and follows the instantaneous frequency curve for different $x_0$.
As the emission gets closer to the horizon, the spot flattens and gets 
strongly redshifted.
\label{fig:gaussian} }
\end{figure*}

Suppose now that the Gaussian coherent state, prepared by the 
inertial observer, is probed by a uniformly accelerated detector
following the worldline 
$\left(a^{-1}\sinh{a\tau}, a^{-1}(\cosh a\tau -1) \right)$
in 1+1D. \Cref{fig:gaussian} shows both the Wigner
function of a Fock and coherent Gaussian states evaluated 
numerically. Each snapshot represents the Wigner function
for a pulse emitted at a different position $x_0$. 
As the intuition would suggest, the closer the emission is 
to the horizon the more redshifted and deformed the wavepacket
is.

A closed analytical form cannot be obtained in this
case but the structure of the Wigner function can be completely 
understood using Gaussian and stationary phase approximation schemes
and its first correction. The detailed treatment is given in 
\cref{sec/approximation-annex}.
For clarity, let's focus on the $\Phi_\alpha(t,x)\Phi_\alpha^*(t',x')$ 
contribution of the Wigner obtained from \cref{eq/Gaussianwavepacket} evaluated 
on the uniformly accelerated trajectory.

The first approximation scheme that we can employ is to approximate the
received
wavepacket by a Gaussian function around its maximum reached at time 
$\tau_r$. Physically, this is the time of reception for the moving
detector. It is obtained by solving $t_\tau - x_\tau + x_0=0$ which gives
the special relativistic result:
\begin{align}
\tau_r = -a^{-1}\ln\left(1 + ax_0\right)
\label{eq/receptiontime-uni-acc}
\,.
\end{align}
Computing the Wigner function is then straightforward and gives
\begin{align}
&W_{\Phi_\alpha\Phi_\alpha^*}(\tau,\omega) = 
	\frac{2p_0}{D_r}
	\exp\left(
		-\frac{D_r^2}{2 \sigma_x^2} (\tau - \tau_r)^2
	\right) \nonumber \\
	&\exp\left(
		-\frac{1}{2}\frac{4\,\sigma_x^2}{D_r^2} 
		\left(\vphantom{\rule{0pt}{2ex}}\omega - \omega_r(1-a (\tau - \tau_r))\right)^2
	\right)
\end{align}
with $D_r = \me^{-a \tau_r}$ the ''gravitational'' redshift and $\omega_r$
the shifted frequency measured by the moving detector,
\begin{align}
\omega_r = p_0\me^{-a \tau_r} = p_0(1 + ax_0)
\,.
\label{eq/receptionfreq-uni-acc}
\end{align}
This is the uniformly accelerated analogue of the Einstein effect.
Thus, we recover directly at this level of approximation the standard results of light
perceived by a uniformly accelerated observer in a special relativistic setting.

While this rough Gaussian approximation allows us to pinpoint the dominant
part of the Wigner function in the time-frequency plane, it is not well suited 
to understand the inner interference pattern. However, the stationary phase
approximation scheme can. Writing the Wigner function as 
$W_{\Phi_\alpha\Phi_\alpha^*}(\tau,\omega) =
	\int_\mathbb{R}
		A(\upsilon ; \tau) \, \me^{\mi \Phi(\upsilon ; \tau,\omega)}
\, \md\upsilon$, 
the stationary phase is meaningful when the velocity of phase oscillations 
is larger then the variations of the modulus. This is indeed the case here since
the phase blows up exponentially compared to the Gaussian decay of the modulus.
Now, we have to find the stationary points and compute the derivatives at those
points. The stationary points $\tau_s$ are solutions of:
\begin{align}
\frac{\partial \Phi}{\partial \upsilon} (\tau_s; \tau,\omega) = 0 \Rightarrow
\left\{
    \begin{array}{ l l }
        \frac{\omega}{p_0}\,\me^{a \tau} = \cosh a\tau_s/2,  & \frac{\omega}{p_0}\,\me^{a \tau}\ge 1 \\
        \emptyset & \mbox{otherwise}
    \end{array}
\right.
\,.
\end{align}
We have two symmetric solutions $\tau_s$ and $-\tau_s$.  The condition
of existence shows that the stationary phase approximation is defined in
the convex hull of the instantaneous frequency curve
$\omega(\tau) = p_0 \, \me^{-a \tau}$.
The second derivative evaluated at $\tau_s$ gives the validity domain of the
stationary phase approximation. On the positive solution $\tau_s$:
\begin{align}
\frac{\partial^2 \Phi}{\partial \upsilon^2} (\tau_s; \tau,\omega) 
	&= -\frac{a }{2} \omega(\tau)
		\sinh\left(\arcosh\left(\frac{\omega}{\omega(\tau)}\right)\right) \nonumber \\
\underset{\tau_s>0}{\Rightarrow}
\frac{\partial^2 \Phi}{\partial \upsilon^2} (\tau_s; \tau,\omega) 
	&= -\frac{a}{2}\sqrt{\omega^2 - \omega^2(\tau)} \le 0
\,.
\end{align}
Away from the instantaneous curve and inside its convex hull, the Wigner function is 
well approximated by the stationary phase approximation. Its explicit derivation is given
in \cref{sec/approximation-annex} and we have:
\begin{align}
W_{\Phi_\alpha\Phi_\alpha^*}(\tau,\omega) &= 
 	\sqrt{\frac{8p_0^2}{a\sigma_x^2}}
 	\frac{\exp\left(
		-\frac{(\omega-\omega_r)^2 + [\omega-\omega(t)][\omega+\omega(t)]}{2(ap_0\sigma_x)^2}
		\right)}{\sqrt{\omega^2 - \omega^2(t)}}
	 \nonumber \\
	&\cos\left( 2\left[\frac{1-\ln2}{a} - \tau\right]\omega + \frac{\pi}{4}\right)
\,.
\end{align}
This form can be interpreted as follows. First, oscillations are present
in the Wigner function, given by the cosine term, covering the whole 
time-frequency plane. Second, the dominant contribution is at the 
intersection of a  strip centered around the frequency $\omega_r$ 
and a tube following the instantaneous frequency $\omega(t)$. 
This allows to understand pictorially why interferences appear as 
we get closer to the horizon: the intersection region gets wider as 
we get closer, allowing the interferences to be visible.

To be rigorous, the approximation fails on the instantaneous curve 
$\omega(\tau)$. We should then go to the next order of approximation:
this is the Airy approximation. Fortunately, since
$\frac{\partial^3 \Phi}{\partial \tau^3} (\tau_s; \tau,\omega(\tau))\ne 0$,
we do not need to go to a higher order. The behavior of the Airy 
function is controlled by the curvature $\epsilon(\tau)$ of the 
instantaneous frequency:
\begin{align}
\epsilon(\tau) 
=\frac{1}{4\pi} \left( \frac{\md^2\omega(\tau)}{\md t^2} \right)^{1/3} 
=\frac{\left( a^2\omega(\tau) \right)^{1/3}}{4\pi} 
\,.
\end{align}
Nonetheless, the stationary phase approximation (and its corrections) 
of the Wigner function gives already the general qualitative
structure of the oscillations that we can see on \cref{fig:gaussian}.

\subsubsection{General $1+1$ d trajectory}
\label{sec/gaussian-general}

\begin{figure*}
	\includegraphics{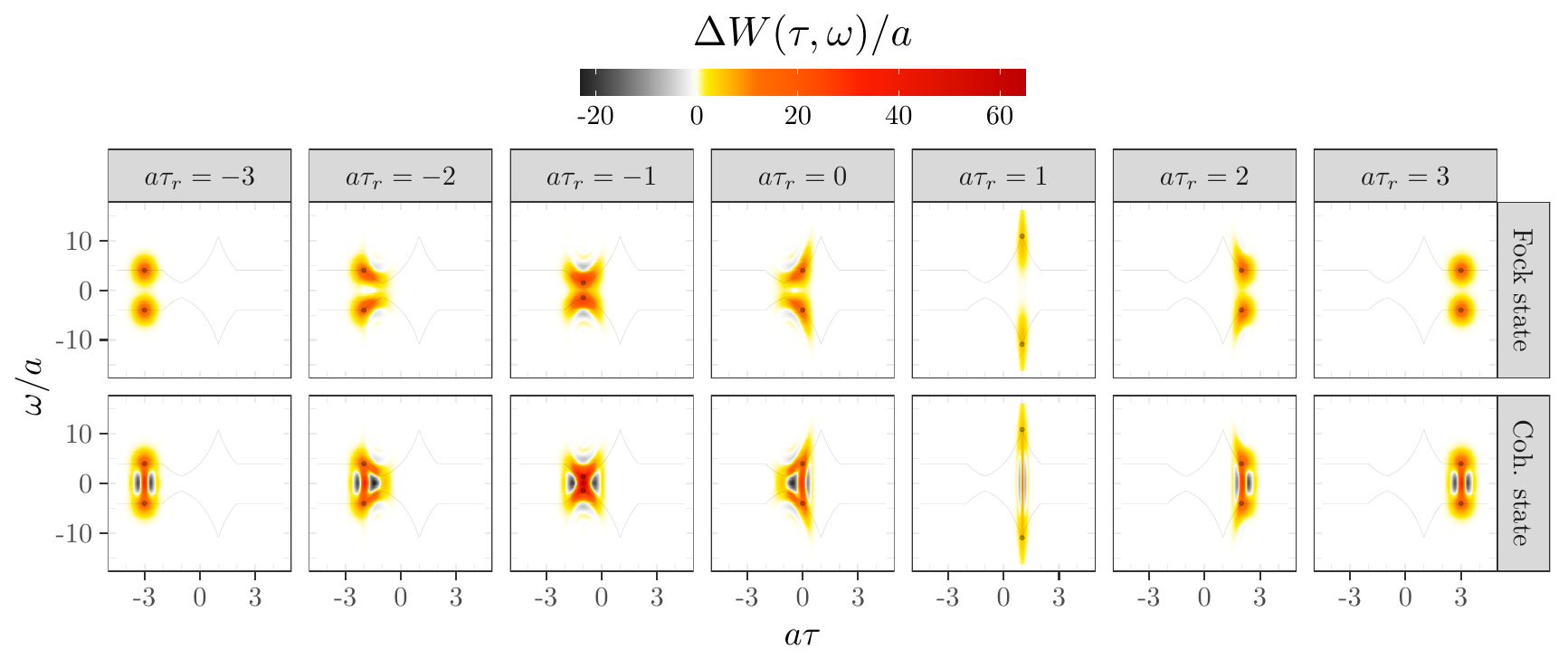}
	\caption{Wigner function representations of a Gaussian
	Fock state (top row) and Gaussian coherent state (bottom row) 
	probed by a detector following a uniformly accelerated twin-like 
	trajectory: the spot is chirped by the accelerated motion with a rate 
	$-a(\tau)\omega(\tau)$ but follows the instantaneous frequency curve.
	The total time of the accelerating phase is $a \tau_{\text{acc}} = 4$, with
	transitions at $\tau = -2, -1, 1, 2$. The frequency of the wavepacket in
	the inertial frame is $p_0/a = 4$ and its width is $a \sigma_x = 1/3$.
	\label{fig/twins-gaussian}}
\end{figure*}

For a generic trajectory, it is of course not possible to obtain a 
complete analytical form of the Wigner function. Nonetheless, its
general features are clearly obtained from the Gaussian and
stationary phase approximations that we already used 
for the uniformly accelerated case. Indeed, from the 
detailed computations presented in \cref{sec/approximation-annex},
we can prove the intuitive idea that first the Gaussian spot 
is shifted in the time-frequency plane by the ''gravitational'' redshift.
The instantaneous frequency curve that the spot is following is 
given by (see \cref{sec/general-motion} for the notations):
\begin{align}
\omega(\tau) = p_0 \, \me^{-A(\tau)}
\text{ with }
A(\tau) = \int_0^\tau a(u) \, \md u
\,.
\end{align}
The spot is centered around the reception time $\tau_r$ which is 
a solution of the equation $\int_0^{\tau_r} \exp(-A(u)) \, \md u = - x_0$.
Since the term in the integrand is positive, this equation possesses
either no or a single solution. This comes from the fact that the
observer necessarily travels slower than the speed of light. As such it
is only possible to meet the photon once. If this equation has no
solution, it means that the photon was emitted behind the event horizon
of the observer. We note that, at this order of approximation, the 
chirp rate is what we classically expect: it is given by the variation 
of the frequency shift for different times which is here 
$\frac{\md \omega(\tau)}{\md \tau} = - a(\tau) \omega(\tau)$.

The inner interference pattern (inside the convex hull defined by the 
instantaneous frequency curve) is once again understood by resorting
to the stationary phase approximation and its Airy correction.

\Cref{fig/twins-gaussian} shows the response of a moving detector
following a uniformly accelerated by parts trajectory. Starting from
inertial motion, the first phase of the motion accelerates uniformly 
with acceleration $a$ at time $a\tau = -2$ up until time $a\tau = -1$. 
The second phase between $a\tau = -1$  and $a\tau =1$ has acceleration 
$-a$. The last phase has again acceleration $a$ up until $a\tau=2$ with
inertial motion onward. This is the kind of trajectory considered
in the twin paradox setup. The signal follows the instantaneous
frequency curve which can be computed exactly in this case and the 
wavepacket is deformed, chirped, along it.

\subsection{Transformation of coherence}

The mathematical framework developed so far is also well suited to analyze
superpositions. Let's again consider a one particle excitation which is now prepared
in a wavepacket $\Phi(x)$ composed  of a linear combination of elementary ones
$\Phi_k(x)$ as:
\begin{align}
	\Phi(x) = \sum_k a_k \Phi_k(x)
\,.
\end{align}
In this case, it is straightforward to show that:
\begin{equation}
	\Delta W(\tau, \omega)
	=
	\sum_{k,k'}
	a_{k'}^* a_{k}
	\Delta W_{k k'}(\tau, \omega)
\end{equation}
where we introduced the notation
\begin{equation}
	\Delta W_{k k'}(\tau, \omega)
	=
	\int_{\mathbb{R}}
	\Phi_{k'}^*(\tau - \upsilon/2)
	\Phi_{k}(\tau + \upsilon/2)
	\, \me^{\mi \omega \upsilon}
	\, \md \upsilon
\,.
\end{equation}
When $k = k'$, we recognize that $\Delta W_{k k'}$ is the excess Wigner
function in the presence of the excitation $\Phi_k$. Furthermore, $k
\neq k'$ indicates cross terms, responsible for the so-called
outer interference terms, between the different components $\Phi_k$. 
Those interferences were already present in 
\cref{fig/twins-gaussian,fig:gaussian} for the Gaussian coherent state.
The total excess Wigner function can thus be expressed as a sum containing
the main components and cross-terms:
\begin{equation}
	\begin{split}
	\Delta W(\tau, \omega)
	&=
	\sum_k |a_k|^2 \, \Delta W_{kk} (\tau, \omega) \\
	&+
	\sum_{k \neq k'} a_{k'}^* a_k \, \Delta W_{k k'} (\tau, \omega)
\,.
	\end{split}
\end{equation}
The important message here is that the excess terms deform naturally. If
we start with some spatial superposition, each term will be deformed as
if it were alone. At the same time, the outer interference terms
depends only on the deformed wavepackets. This means that if we emit a
wavepacket in a linear superposition of two wavepackets received around
times $\tau_1$ and $\tau_2$, we expect that the interference terms will be
located at the midpoint $\tau_m = (\tau_1+\tau_2)/2$. Furthermore, those
interference terms will not depend on the details of the trajectory
at time $\tau_m$ but on those at the times of reception $\tau_1$
and $\tau_2$.

\begin{figure}
	\includegraphics{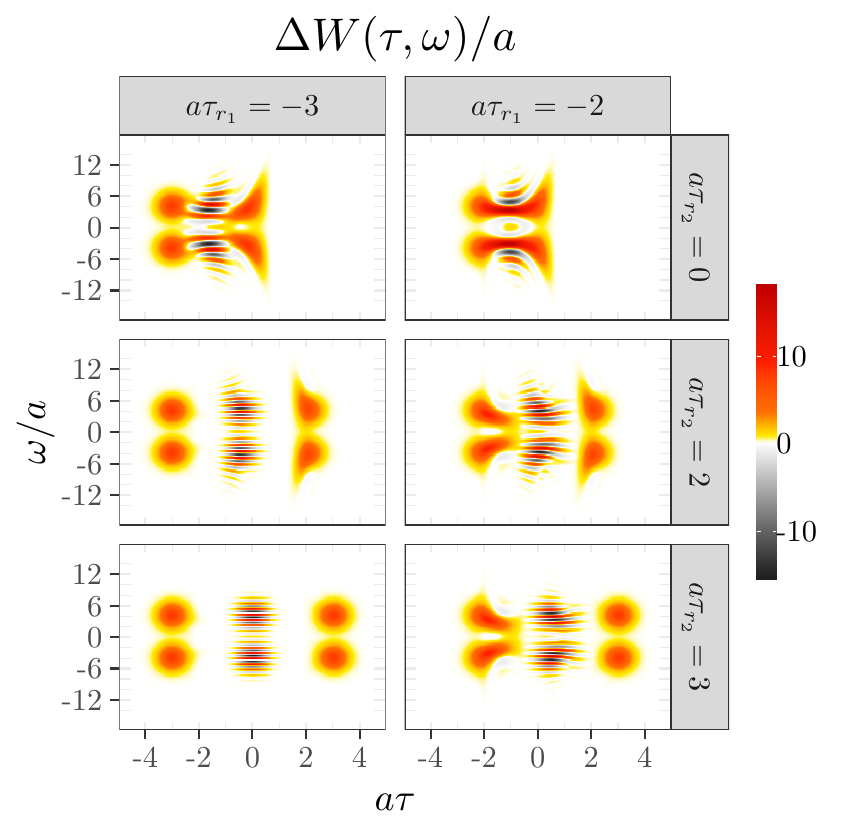}
	\caption{Wigner function representation of a superposition of a Gaussian
	photon wavepacket probed by a detector following a uniformly accelerated 
	twin-like trajectory for different $\tau_{r,1}$ and $\tau_{r,2}$ 
	reception times respectively associated to the first and second 
	component of the superposition.
	The total time of the accelerating phase is $a \tau_{\text{acc}} = 4$ with
	transitions at $\tau = -2, -1, 1, 2$. The wavepacket is emitted at
	frequency $p_0/a = 4$ in the inertial frame with a width $a \sigma_x =
	1/3$.
	\label{fig/twins-superp}}
\end{figure}

\Cref{fig/twins-superp} considers once again the uniformly accelerated
twin-like trajectory of \cref{sec/gaussian-general}. The field is however
prepared in a spatial superposition of two Gaussian wavepackets (only the
photon wavefunction is represented for clarity):
\begin{align}
\Phi(x) = \Phi_2(x) + \Phi_1(x)
\end{align}
where the wavepacket $\Phi_i(x)$ is centered around the position
$x_i$ and is received by an inertial observer at time $t_{r_i}$ and by 
the moving detector at times $\tau_{r_i}$. Quite naturally, the 
structure of the Wigner function depends on the spatial separation 
of the components of the superposition or, equivalently, on the 
detection times, and the local characteristics of the trajectory.

First, by denoting $\mathcal{J}^-(x)$ the causal past of a point $x$
in spacetime, the coherence properties are modified by the motion of the
detector only if at least one component has been prepared in the 
spacetime region $\mathcal{J}^-(f) \setminus \mathcal{J}^-(i)$ where 
$i$ and $f$ are respectively the beginning and end events of the acceleration
phase.

The second feature concerns the delay time between the reception 
of the two wavepackets. On \cref{fig/twins-superp}, the wavepackets
were prepared such that the inertial delay 
$\Delta t_r = 4 a^{-1} \sinh(a \Delta\tau_r /4) $ with $\Delta\tau_r = 3$
which is the general twin-paradox delay formula for this trajectory. 
This time-delay is clearly seen in the Wigner function and satisfies the 
special relativistic result (\cref{fig/comp-twin-super} compares the 
inertial and accelerated responses directly). 

Finally, while the coherence pattern is identical to a pure inertial 
response when the packets are prepared outside the region 
$\mathcal{J}^-(f) \setminus \mathcal{J}^-(i)$, the interference
pattern is clearly deformed by the motion of the detector when 
one component is probed in the accelerated phase.

\begin{figure}
	\centering
	\includegraphics{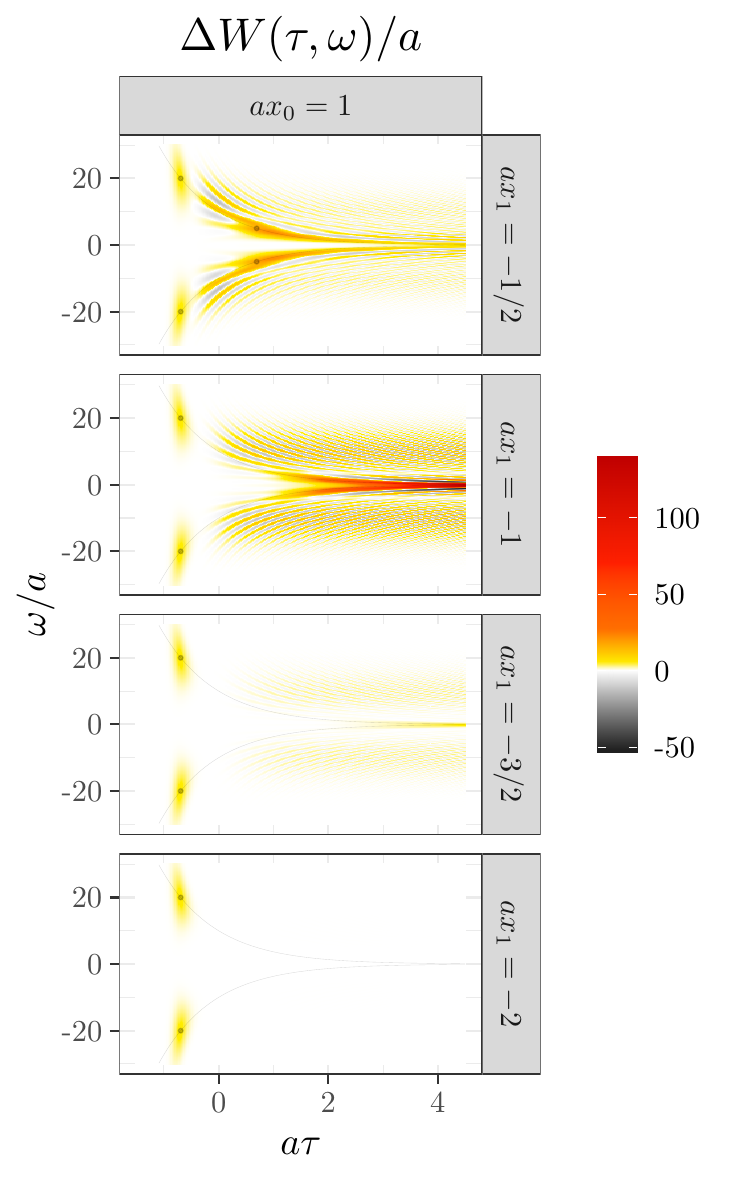}
	\caption{Wigner function representations of a Gaussian superposition received
	at different times for a uniformly accelerated observer. The coherences are spread
	and eventually lost when a member of the superposition gets close or cross
	the horizon. The wavepacket is emitted at frequency $p_0/a = 10$ in the
	inertial frame with a width $a \sigma_x = 1/5$.
	\label{fig/gaussian-super-uniacc}}
\end{figure}

\Cref{fig/gaussian-super-uniacc} shows the more extreme case
of the evolution of coherence of a Gaussian superposition probed
by a uniformly accelerated detector. The spacetime geometry
probed by this detector, also called Rindler spacetime or wedge, 
is quite different than the previous case because of the presence
of an event horizon. Naturally, a detection event occurs if and only 
if the wavefunction has been  prepared with a support in the wedge.
The situation shown in \cref{fig/gaussian-super-uniacc} represents 
a Gaussian superposition of two wavepackets, one of which propagates
closer and closer to the horizon. The coherence gets spread and 
redshited as the wavepacket approaches the horizon which is a 
consequence of the same effects happening to the wavepacket 
itself. 

After crossing the horizon, no detection signals can be
recovered by the detector and the coherences are lost. This is 
of course the same effect happening in black hole physics which
leads to the famous information paradox.
We should note that this loss of coherence cannot be properly 
qualified as a decoherence process in the traditional sense where
an environment interacts with the system and attenuates the 
interference pattern. Indeed, the deformation and loss of 
coherence only comes about because the wavepacket themselves
are deformed by motion or lost behind an horizon which is not 
what decoherence is about.

\section{Discussions} 
\label{sec/bigpicture}

So far, the point of view we adopted was a pure signal processing one.
Indeed, our interest was only focused on doing a proper analysis of 
signals characterized by correlation functions 
$G_\rho(\tau_1,\dots,\tau_n)$ obtained from a set of point-like
detectors. The question remains on how to relate those signals to 
quantum field theories for different observers 
\cite{Padmanabhan-1982,Grove-1983,Padmanabhan-2002}. 

The fundamental question at this stage is to understand what can 
be reconstructed about the quantum field from the signals (of one 
or a set of detectors). To be more precise, we can roughly group the
questions in two categories:
\begin{itemize}
\item What can we learn about the trajectory of the detector with respect to 
laboratory frame $\theta$?
\item Can we reconstruct a field theoretic picture 
$(\mathcal{H}_\theta, f_\theta)$ from the signals, with
$\mathcal{H}_\theta$ the Hilbert space of the theory and 
$f_\theta$ the mode on which the field is decomposed?
\end{itemize}
Our signal processing approach opens up some interesting 
perspectives on those questions. 

Concerning the recovery of information about the trajectory, we need 
a prior information about what was sent by the laboratory. Indeed, it is 
conceivable to imagine only inertial detectors probing a state prepared 
in such a way as to simulate an accelerated response. So to not be fooled 
by what we measure, we need, for instance, the laboratory to communicate
to the accelerated observer what they originally prepared. Information
about the trajectory can then be recovered by properly fitting the 
measured signal or, if we have enough data, reconstruct the instantaneous
frequency curve from which the acceleration can be deduce since
$\md\omega(\tau)/\md \tau =-a(\tau) \omega(\tau)$. If no excitations
are present, we can still have some information about the trajectory
from the power spectrum \cref{eq/power} of the vacuum: indeed, 
for a one dimensional motion, it is directly proportional to the square
of the acceleration.

The second question was about reconstructing a field theoretic
or many-body point of view from the signals. While we are not going 
to investigate thouroughly this complicated question, time-frequency 
analysis can shed some light on one particular issue concerning the 
definition of a notion of particles. 

In the standard many-body approach, there is no issue to define a notion
of particles in a stationary situation~\cite{Ashtekar-1975} like a
uniformly accelerated motion. Qualitatively, we have a notion of time 
from which we can define a Fourier transform. There is however no general
method to define a notion of particles in non-stationary situations.
In other words, the notion of particles is an emerging notion~\cite{Haag-1996}.
It is nonetheless interesting to link this emerging notion to the notion
encountered in the standard many-body approach.

One way to do this is to introduce an operational notion of
particles thanks to response signals $G(\tau_1,\tau_2)$ of 
detectors~\cite{Unruh-1976}. The question is to then relate those two
notions which, in general, are quite different.
As we already mentioned, it is valid to interpret 
$G_\rho(\omega,\omega')$ in terms of excitations for inertial detectors
as is usually done, for instance, in quantum optics: the two notions
coincide. This breaks down \emph{a priori} in non-stationary 
motions. 

The time-frequency analysis of the complete signal offers a strategy
to link the two notions and reconstruct a many-body particle
interpretation. Indeed, from the full signal, it is possible to extract
stationary domains~\cite{Flandrin-2010}. Intuitively speaking, we 
can extract domains where it is meaningful to decompose the 
field modes like $f_\mathbf{p}(\tau, \mathbf{r}) \propto 
\me^{-\mi \omega_\mathbf{p} \tau} f_\mathbf{p}(\mathbf{r})$
\cite{Grove-1983}. Knowing then the stationary time scales and 
averaging the signal over them, notions of particles could then be 
locally defined. Operationally speaking, 
what can be done is to consider a detector with a response 
function having support on those domains.

To illustrate this strategy, let's consider again the situation
studied in \cref{sec/adiabatic} where we considered an oscillatory 
motion of the form ${a(\tau) = a_0 + a_1 \sin(2 \pi f \tau)}$
represented in \cref{fig/sinus}. This situation is completely 
non stationary and even not globally adiabatic when $a_0 = a_1$.
No natural particle interpretation can be found. Nonetheless, 
we know that we can consider the signal as approximately stationary
around a given time $\tau$ over a timescale $\tau_{\text{s}}$ (depending
itself on $\tau$): this is the same condition controlling the validity 
of the functional expansion \cref{eq/functional-expansion}.

\begin{figure}
	\begin{center}
		\includegraphics{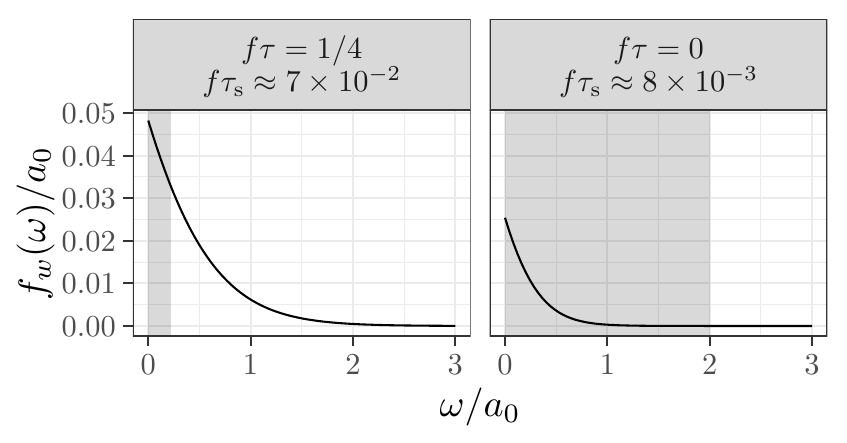}
	\end{center}
	\caption{
		Gaussian average of the Wigner function over a window given
		by $\tau_{\text{s}}$ such that $|a/\delta a| \le 1/20$ for the
		sine case where $a_0 = a_1$, and $2 \pi f/a_0 = 1/5$. To have
		a consistent time-frequency representation, we need to perform
		another Gaussian average of typical spread given by the shaded
		area which depends on $\tau_{\text{s}}$. A particle picture is 
		meaningful only above this frequency threshold.
	\label{fig/particles}}
\end{figure}

In a given signal, different stationary time scales exist as we
discussed already in \cref{sec/breakdown}. \Cref{fig/particles}
represents the (Gaussian) average of the Wigner function around
a time $\tau$ with a time window of typical width $\tau_{\text{s}}$: 
only a section is represented (it is sufficient since we are
approximately stationary) and corresponds to the averaged energy 
distribution $f_w(\omega)$. 
To be consistent and respect the Heisenberg time-frequency
indeterminacy, a (Gaussian) average should be performed in frequency:
this frequency window is the orange area in \cref{fig/particles} where
two extreme cases are shown :
\begin{itemize}
\item At the maximum of acceleration (where the adiabatic regime is valid,
$f\tau = 1/4$), the distribution $f_w(\omega)$ is a thermal distribution 
at a temperature around $a_\tau$. The frequency average is such 
that the overall time-frequency response of a detector with a
response function of width $\tau_{\text{s}}$ around $f\tau = 1/4$
will be a stationary thermal distribution. We can then reconstruct in
this time-frequency domain a uniformly accelerated particle picture.
\item At lower accelerations, the distribution $f_w(\omega)$
looks thermal but has to be averaged over a very large frequency domain
compared to its typical width. This is a consequence of the very low
stationary timescale there. In this case, the only meaningful particle 
picture that can be constructed is at very high frequencies 
and matches one of an inertial response. This follows the intuition that
high-frequency modes are equal to inertial modes.
\end{itemize}

In the end, the main lesson from this discussion is that particles
emerge from the signal and applying a time-frequency analysis seems the most
appropriate to tackle the issue of particle reconstruction and 
to link the operational and many-body definitions.

\section{Conclusion} 
\label{sec/conclusion}

In this paper, we introduced a signal processing time-frequency approach
to the problem of detectors in motion in relativistic quantum field
theory. It offers a natural and synthetic framework to analyze 
non-stationary trajectories. We provided a detailed analysis of the 
adiabatic regime, its corrections and its breakdown. We then moved 
on to study how excitations are probed by a moving detector, 
focusing for clarity on Gaussian states. The structure of the Wigner
function can be completely understood using simple approximation
schemes. Beside recovering time-frequency special relativistic behaviors
in general frames, we were able to analyze how wavepackets and 
their coherence properties are transformed by the motion of the 
detector.

We finally used our analysis of excitation and motion to discuss how
time-frequency analysis provides a promising approach to 
clarify conceptual questions behind the problem of moving detectors,
especially concerning the definitions of  a notion of particles.
Indeed, time-frequency allows to define a notion of relative
stationary timescales over the signal, permitting than to locally
link the operational and many-body definitions of particles.

Apart from those conceptual questions, this time-frequency
signal processing approach opens up many interesting perspectives
to sharpen our understanding of relativistic detectors. 
Indeed, a natural generalization is to analyze higher order correlation 
functions, which play an important role in quantum optics. This would 
allow to understand from first principles the interplay between 
entanglement \cite{Fuentes-2005,Ralph-2019}, which is encoded in the second order correlation
function, and motion in a completely relativistic setting.
Moreover, one of the main challenge is to have an experimental access 
to the Wigner function. Measuring the Wigner is traditionally 
done through interferometric setup like the Hong-Ou-Mandel
experiment. This then demands to properly analyze 
those interferometric experiments when probed by a moving 
detector.
Finally, the same approach could be generalized to curved 
spacetime, allowing again to understand the response of detector
in non stationary spacetime situations like the formation of black 
holes by a collapsing star or the effect of gravitation on the 
entanglement of quantum systems.

\acknowledgements{%
	We thank P.~Degiovanni and G.~Fève for useful discussions. We also thank
	J.-M.~Raimond for useful remarks on our manuscript.
}

%% file: Wigner_in_relativistic_QFT_app.tex
\section{Wigner function of a discontinuous acceleration}
\label{sec/discontinuity}

We will consider the case where an observer is going from an inertial
phase to a uniformly accelerated phase at acceleration $a$. Its
trajectory in its proper time can be written as
\begin{equation}
	x(\tau) =
	\begin{cases}
		(\tau, 0) & \text{if $\tau \le 0$,} \\
		\left(a^{-1} \sinh a \tau, a^{-1}(\cosh a \tau -1)\right) & \text{if $\tau
		\ge 0$
		\,.}
	\end{cases}
\end{equation}
We see that this expression has a discontinuity in the second derivative
of $x(\tau)$. We expect the expression of $G(\tau +
\upsilon/2, \tau - \upsilon/2)$ to have discontinuities in second- or higher-order
derivatives. Since algebraic high-frequency behaviors of the Fourier
transform are determined by those discontinuities, we will take some
care to analyze them. For this we will rewrite
\begin{equation}
	G(\tau + \upsilon/2, \tau - \upsilon/2)
	=
	f_\tau(\upsilon) + g_\tau(\upsilon)
\,,
\end{equation}
where $f$ contains the lower-order discontinuities of $G$, and
$g$ may contain higher-order discontinuities. While the function $f$ is
arbitrary, its high-frequency behavior only depends on the
discontinuities, and not on the precise details of $f$.

In this case, $G$ has discontinuities at $\upsilon = \pm 2 \tau$. 
Since $G(\tau + \upsilon/2, \tau - \upsilon/2)$ is even in
$\upsilon$, we will consider only the case $\upsilon \ge 0$. We will
choose for $f$ an even truncated polynomial, to take into account this
symmetry
\begin{equation}
	f_{\tau}(\upsilon)
	=
	\frac{\alpha}{n!} (\upsilon - 2 \tau)^n (\upsilon + 2 \tau)^n
	\Pi_{[-2\tau, 2\tau]}(\upsilon)
\,,
\end{equation}
where $n$ is the order of the discontinuity, $\Pi$ is the gate function
and $\alpha$ is such that $f$ captures the discontinuities of $G$,
\begin{equation}
	\partial_\upsilon^n G\rvert_{\upsilon = 2 |\tau|^+}
	-
	\partial_\upsilon^n G\rvert_{\upsilon = 2|\tau|^-}
	=
	\partial_\upsilon^n f_{\tau}\rvert_{\upsilon = 2 |\tau|^+}
	-
	\partial_\upsilon^n f_{\tau}\rvert_{\upsilon = 2 |\tau|^-}
\,.
\end{equation}

For $\tau \ge 0$, the discontinuity happens in the second derivative of
$G$. After performing the Fourier transform of the corresponding
$f_\tau$, we find that the high-frequency behavior is expressed as
\begin{equation}
	\Delta W(\tau, \omega) \simeq
	- \frac{1}{4\pi^2} \frac{a^4}{8 \sinh^2 a \tau} \frac{\sin 2 \omega \tau}{\omega^3}
\,.
\end{equation}
For $\tau \le 0$, $G$ has a discontinuity in its third derivative. The
same technique gives the result
\begin{equation}
	\Delta W(\tau, \omega) \simeq
	- \frac{1}{4\pi^2} \frac{a^2}{16 \tau^3} \frac{\cos 2 \omega \tau}{\omega^4}
\,.
\end{equation}

\section{Adiabatic regime}
\label{sec/adiab-annex}

\begin{figure*}
	\begin{center}
		\includegraphics{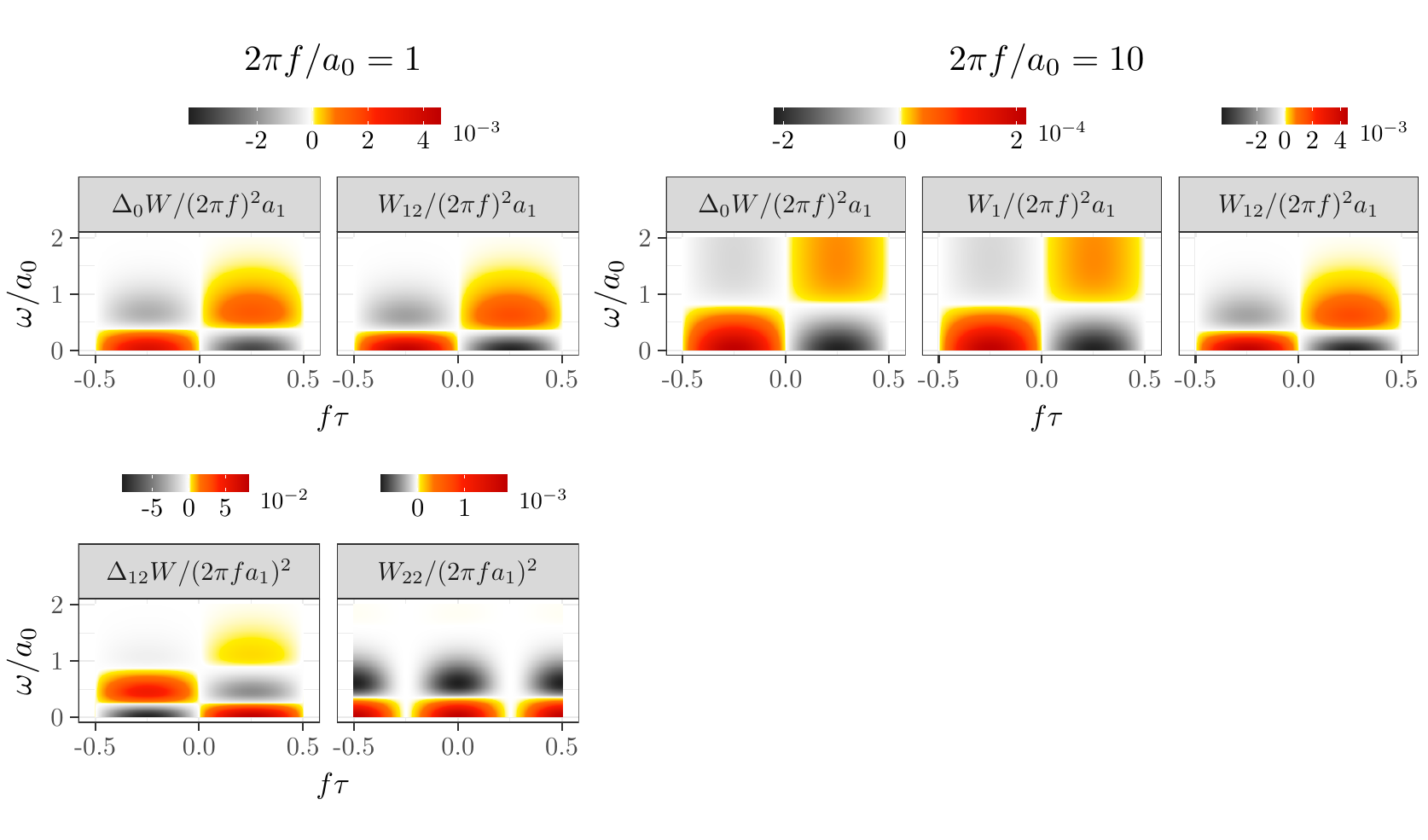}
	\end{center}
	\caption{
		Different breakdowns of adiabatic expansions when the first
		derivative $\dot{a}$ is small for a sinusoidal acceleration.
		In these plots we keep $2 \pi f a_1/a_0 = \num{e-2}$. On the
		left, the frequency being of the order of $a_0$, the adiabatic
		development breaks down after the correction in $\ddot{a}$. On
		the right, the frequency is much higher than $a_0$. While the
		perturbative expansion works perfectly well, the adiabatic
		development breaks down, even at first order. We see also that
		corrections are important outside of the thermal bandwidth in
		this case.
	}
	\label{fig/adia-expand-break}
\end{figure*}

In this appendix, we will give some further details about the adiabatic
development. We will start by the functional development, up to first
order in \cref{sec/adiab/funcdev}. We will then focus on the derivative
development in \cref{sec/adiab/derivdev}. The validity of the different
developments can be seen on \cref{fig/adia-expand-break}.

\subsection{Functional development}
\label{sec/adiab/funcdev}

In order to perform the derivative developmment, we will write the
acceleration as
\begin{equation}
	a(\tau + \upsilon) \simeq a(\tau) + \delta_\tau a(\upsilon)\,.
\end{equation}
We will see that, provided that $\delta_\tau a(\upsilon) \ll a(\tau)$
for $|\upsilon| < \tau_{\text{s}}$ and $a \tau_{\text{s}} \gg 1$, this
development is valid. We can rewrite the first order term as
\begin{equation}
	W_1(\tau,\omega)
	=
	\frac{1}{4 \pi^2}
	\int \frac{a(\tau)^3}{4 \sinh^3 (a(\tau) \upsilon / 2)}
	\int_{-\upsilon/2}^{\upsilon/2}
	\sinh(a(\tau) \upsilon') \, \delta_{\tau} A(\upsilon') \, \md \upsilon'
	\ 
	\me^{\mi \omega \upsilon} \, \md \upsilon\,,
\end{equation}
with $\delta_\tau A(\upsilon) = \int_{0}^{\upsilon} \delta_{\tau} a(\upsilon') \md
\upsilon'$.

From this development, we can see that the first-order term only depends
on the odd part of $\delta_\tau A$, and thus the even part of
$\delta_\tau a$. Furthermore, if $\delta_\tau A$ has a polynomial
behavior, the integrand has an exponential cut-off in $|\upsilon|$.
Provided that $\delta_\tau a$ is small enough compared to $a(\tau)$ over
a scale $\tau_{\text{s}} \gg a(\tau)^{-1}$, this first-order correction
will be small compared to the thermal behavior.

Since this term is a linear functional in $\delta_{\tau} a$, we can
treat it through Fourier analysis. Since we are only sensitive to the
even part of $\delta_\tau a$, we can set
\begin{equation}
	\delta_{\tau} a(\upsilon)
	=
	a_1(\tau) (\cos(2 \pi f \upsilon) - 1)\,.
\end{equation}

In this case, we find that
\begin{equation}
W_1
=
\frac{a_1}{4\pi^2}
\Big[
\frac{1}{1+(2\pi f /a_\tau)^2}\frac{g_+ + g_-}{2} -
\frac{\omega/2\pi f}{1+(2\pi f /a_\tau)^2}(g_+ - g_-) +
\frac{2\pi}{a_\tau}\omega \dot{g}_0-g_0
\Big]\,.
\end{equation}

If $2 \pi f \ll a_\tau$, we can use a perturbative development
\begin{equation}
	W_1(\tau, \omega)
	=
	\frac{a_1}{4 \pi^2} \left(\frac{2 \pi f}{a(\tau)}\right)^2
	\left[
		-1 + \frac{2 \pi}{a(\tau)} \omega \partial_x
		+ \frac{1}{2} (\pi \partial_x)^2
		- \frac{1}{3} \frac{\omega}{a(\tau)} (\pi \partial_x)^3
	\right]g\left(\frac{2 \pi}{a(\tau)} \omega \right)\,,
\end{equation}
with $g(x) = x/(\exp(x)-1)$.

This correction is vanishing quadratically as frequency is lowered.
Furthermore, all the corrective terms act in the thermal bandwidth.
The first two terms can be seen as a correction to the thermal state by
a difference of temperature $-a_0 (2 \pi f/a(\tau))^2$. This correction
can be interpreted as a averaging effect due to the even part of the
acceleration.

On the contrary, when $2 \pi f \gg a(\tau)$, we can approximate the
corrective term with
\begin{equation}
	W_1(\tau, \omega)
	=
	\frac{a_1}{4 \pi^2}
	\left[
		-g_0 + \frac{2 \pi}{a_1} \omega g_0'
	\right]
	+
	\frac{a_1}{4 \pi^2} \frac{g_-}{1 + (2 \pi f/a(\tau))^2}
	\left( \frac{1}{2} + \frac{\omega}{2 \pi f} \right)\,.
\end{equation}
The first term corresponds to a shift of the temperature to the average
value of $\delta_{\tau} a$. At high frequencies, the temperature is thus
blurred by the average behavior.
The second term is exponentially small for frequencies higher than $\pi
f$. Below $\pi f$, it behaves like the quadratic form:
\begin{equation}
	\frac{a_0}{4 \pi^2} \frac{a(\tau)}{4 f}\left( 1 -
	\left(\frac{\omega}{\pi f}\right)^2 \right)\,.
\end{equation}
Thus, this correction to the average thermal behavior has a bandwidth
much larger than the one of thermal fluctuations.

\subsection{Derivative development}
\label{sec/adiab/derivdev}

We saw in the previous development that the linear term in $\delta_\tau
a$ only depends on the even part of $\delta_\tau a$. Since in
the adiabatic regime the frequencies contained in $a(\tau)$ around a
window of size $\tau_\text{s}$ are expected to be sufficiently low
compared to the $a(\tau)$, we should have the splitting of order $\dot
a(\tau) \gg \ddot a(\tau)$, and higher order derivative should be
negligible in front of those terms. 
As such, we will develop to the terms $\ddot a(\tau)/a(\tau)^3$ and $\dot
a(\tau)^2/a(\tau)^4$.

The term in $\ddot a(\tau)/a(\tau)^3$ reads
\begin{equation}
	W_{12}(\tau, \omega)
	=
	-\frac{1}{4 \pi^2} \frac{\ddot{a}(\tau)}{a(\tau)^2}
	\left[
		-1 + \frac{2 \pi}{a(\tau)} \omega \partial_x
		+ \frac{1}{2} (\pi \partial_x)^2
		- \frac{1}{3} \frac{\omega}{a(\tau)} (\pi \partial_x)^3
	\right]g\left(\frac{2 \pi}{a(\tau)} \omega \right)\,.
\end{equation}
We recover here the approximation of the linear term in case of small
frequencies.

The terms in $\dot{a}(\tau)^2/a(\tau)^4$ reads
\begin{equation}
	W_{22}(\tau,\omega)
	=
	\frac{2}{3 \pi^2}
	\frac{\dot{a}(\tau)^2}{a(\tau)^3}
	\left[
		-1 + \frac{2 \pi}{a(\tau)} \omega \partial_x
		+ \left(
			\left(\frac{\omega}{2 a(\tau)}\right)^2
			+ \frac{5}{8}
		\right) (\pi \partial_x)^2
	\right] g\left(\frac{2 \pi}{a(\tau)} \omega\right)\,.
\end{equation}
In this case, the first two terms can also be seen as a small thermal
shift by $8\dot{a}^2 / 3 a^3$.

\section{Fock and mono-chromatic coherent state}
\label{sec/fock-coherent}

In this appendix, we present the computation of the Wigner function
of a mono-chromatic coherent state and Fock state.
First let's suppose we have a single mode coherent state 
$\ket{\alpha_{\mathbf{p}}}$. Given the commutation relation that 
we have, the overlap between two coherent states is given by 
\begin{align}
\langle \alpha_{\mathbf{p}}|\alpha_{\mathbf{p'}}\rangle = 
2\omega_{\mathbf{p}} (2\pi)^3 \delta(\mathbf{p}-\mathbf{p'})
\,.
\end{align}
Then the first order coherent function can be computed directly has 
\begin{align}
\braket{\alpha_{\mathbf{p}} |\phi(\tau)\phi(\tau')| \alpha_{\mathbf{p}}}
=
\braket{ 0|\phi(\tau)\phi(\tau')| 0 }
+
\alpha_{\mathbf{p}}^2 
\me^{-\mi \left[ \omega_\mathbf{p} (t+t') 
- \mathbf{p}.(\mathbf{x}+ \mathbf{x'}) \right]}
+
\alpha_{\mathbf{p}}\overline{\alpha}_{\mathbf{p}}
\me^{\mi \left[ \omega_\mathbf{p} (t-t') 
- \mathbf{p}.(\mathbf{x}- \mathbf{x'}) \right]}
+
\text{ h.c.}
\label{first-correlation-coherent}
\end{align}
where $(t,\mathbf{x})$ are functions of the proper time $\tau$.

Let's check quickly what we obtain for inertial trajectories. For one 
with zero velocity $(\tau,0)$, we have 
\begin{align}
W_{\ket{\alpha_{\mathbf{p}}}}(\tau,\omega) = 
W_{\ket{0}}(\tau,\omega) +
|\alpha_{\mathbf{p}}|^2 
\big( \delta(\omega+\omega_\mathbf{p}) +
\delta(\omega-\omega_\mathbf{p}) \big) +
2\Re\left(
\alpha^2_\mathbf{p} \me^{-2 \mi \omega_\mathbf{p}\tau} 
\right)
\delta(\omega)
\,.
\end{align}
For an inertial observer with velocity $\mathbf{v}$ which then 
has the trajectory $(\gamma \tau, \gamma v \tau)$, the Wigner 
function remains the same except that the frequency of the coherent
state is Doppler shifted to $\omega^{\mathbf{v}}_\mathbf{p}
=\gamma \left[\omega_\mathbf{p} - \mathbf{v}.\mathbf{p} \right]$.

Now consider the uniformly accelerated trajectory
$\left(a^{-1}\sinh{a\tau}, a^{-1}(\cosh a\tau -1) \right)$. 
From \eqref{first-correlation-coherent} we have to compute two 
different integrals:
\begin{align}
W^a_{\ket{\alpha_{\mathbf{p}}}}(\tau,\omega) = 
W^a_{\ket{0}}(\tau,\omega) +
\alpha_{\mathbf{p}}^2 \, I^+_1(\tau,\omega) + 
\alpha_{\mathbf{p}} \overline{\alpha}_{\mathbf{p}}\,
\big( I^+_2 (\tau,\omega) + I^-_2 (t,\omega) \big) +
 \overline{\alpha}_{\mathbf{p}}^2 \, I^-_1(\tau,\omega) 
\end{align}
with
\begin{subequations}
\begin{align}
I^+_1 (\tau,\omega) &= \me^{-2\mi p}  \int_{\mathbb{R}}
\me^{-2\mi 
\left[ \omega_p a^{-1}\sinh a \tau -
p_x a^{-1}\cosh a\tau
\right]
\cosh a\upsilon/2 }
\me^{\mi \omega \upsilon}
\,
\md \upsilon \\
I^+_2 (\tau,\omega) &= \int_{\mathbb{R}}
\me^{2\mi 
\left[ \omega_p a^{-1}\cosh a \tau -
p_x a^{-1}\sinh a \tau
\right]
\sinh a\upsilon/2 }
\me^{\mi \omega \upsilon}
\,
\md \upsilon
\,.
\end{align}
\end{subequations}
Since $\omega_p = p$, we can say that the function 
$f(\tau) = 2 \left[ \omega_p a^{-1}\cosh a \tau \mp
p a^{-1}\sinh a \tau \right]$ is always positive and independent of 
$\upsilon$. Here we choose the coherent state momentum  to be 
in the same direction as the accelerated observer, $p_x>0$. 
The case with opposite momentum is treated in the same way.
Furthermore, the phase in front of the integral of $I^+$ can be 
absorbed by a change of phase of the coherent state and will 
then be omitted in the following.
Thus we have to compute 
\begin{subequations}
\begin{align}
I_1^\pm (\tau,\omega) = \int_{\mathbb{R}}
\me^{\pm \mi  f(\tau)
\cosh a\upsilon/2 }
\me^{\mi \omega \upsilon}
\,
\md \upsilon
\\
I_2^\pm (\tau,\omega) = \int_{\mathbb{R}}
\me^{\pm\mi f(\tau) 
\sinh a\upsilon/2 }
\me^{\mi \omega \upsilon}
\,
\md \upsilon
\,.
\end{align}
\end{subequations}
The first integrals corresponds exactly to an integral representation of the 
Hankel functions defined as 
\begin{subequations}
\begin{align}
H^{(1)}_\nu (x) = \frac{\me^{-\frac{\nu \pi \mi}{2}}}{\mi\pi}
\int_{\mathbb{R}} \me^{\mi x \cosh t -\nu t} \, \md t
\text{ for } x>0, |\Re(\nu)| <1 \\
H^{(2)}_\nu (x) = - \frac{\me^{\frac{\nu \pi \mi}{2}}}{\mi\pi}
\int_{\mathbb{R}} \me^{-\mi x \cosh t -\nu t} \, \md t
\text{ for } x>0, |\Re(\nu)| <1 
\,.
\end{align}
\end{subequations}
So we directly have 
\begin{subequations}
\begin{align}
I^+_1(\tau,\omega) &= \frac{2 \mi \pi}{a}  \me^{\frac{\omega \pi}{a}} 
			H^{(1)}_{-2\mi \omega/a} \big(f(\tau)\big) 
			=\frac{4}{a} K_{-2\mi \omega/a}\big(-\mi f(\tau)\big) 
 \\
I^-_1(\tau,\omega) &= -\frac{2 \mi \pi}{a}  \me^{-\frac{\omega \pi}{a}} 
			H^{(2)}_{-2\mi \omega/a} \big(f(\tau)\big) 
			=\frac{4}{a} K_{2\mi \omega/a}\big(\mi f(\tau)\big) 
\end{align}
\end{subequations}
where the special functions $K_\nu$ are modified Bessel functions of the 
second kind. In short 
\begin{align}
I_1^\pm (\tau,\omega) =\frac{4}{a} K_{\mp 2\mi \omega/a}
\big( \mp \mi f (\tau) \big) 
\,.
\end{align}

The second integral is also related to Hankel functions or Bessel functions.
By deforming the contour of integration by a translation of $\pm\mi\pi/2$, we have 
\begin{align}
\int_{\mathbb{R}} \me^{\mi x \sinh t + \mi\omega t} \, \md t
=
\int_{\mathbb{R}} \me^{\mi x \sinh (t + \mi \pi/2) + \mi\omega (t + \mi \pi/2)} 
\, \md t
= 
\me^{-\frac{\omega\pi}{2}}
\int_{\mathbb{R}} \me^{- x \cosh t + \mi\omega t} \, \md t
= 2
\me^{-\frac{\omega\pi}{2}}
K_{\mi \omega}(x)
\end{align}
so that 
\begin{align}
I_2^\pm (\tau,\omega) 
= \frac{4}{a} \me^{\mp\frac{\omega\pi}{a}} K_{2 \mi \omega /a}(f(\tau))
\,.
\label{eq/wigner-bessel}
\end{align}

The case of Fock states are treated in the same way. In fact, all computations 
have already been done in the case of coherent states, see \eqref{eq/wigner-bessel}. 
For a single mode Fock state
\begin{align}
\braket{n_{\mathbf{p}} |\phi(\tau)\phi(\tau')| n_{\mathbf{p}}}
=
\braket{ 0|\phi(\tau)\phi(\tau')| 0 }
+
n_{\mathbf{p}}
\Big(
\me^{\mi \left[ \omega_\mathbf{p} (t-t') 
- \mathbf{p}.(\mathbf{x}- \mathbf{x'}) \right]}
+ 
\me^{-\mi \left[ \omega_\mathbf{p} (t-t') 
- \mathbf{p}.(\mathbf{x}- \mathbf{x'}) \right]}
\Big)
\,.
\label{first-correlation-fock}
\end{align}
We can check directly that 
\begin{subequations}
\begin{align}
W_{\ket{n_{\mathbf{p}}}}(\tau,\omega) &= 
W_{\ket{0}}(\tau,\omega) +
n_{\mathbf{p}} 
\big( \delta(\omega+\omega_\mathbf{p}) +
\delta(\omega-\omega_\mathbf{p}) \big) \\
W^\mathbf{v}_{\ket{n_{\mathbf{p}}}}(\tau,\omega) &= 
W_{\ket{0}}(\tau,\omega) +
n_{\mathbf{p}} 
\big( \delta(\omega+\omega^\mathbf{v}_\mathbf{p}) +
\delta(\omega-\omega^\mathbf{v}_\mathbf{p}) \big) \\
W^a_{\ket{n_\mathbf{p}}}(\tau,\omega) &= 
W^a_{\ket{0}}(\tau,\omega) +
n_\mathbf{p}\,
\big( I^+_2 (t,\omega) + I^-_2 (t,\omega) \big) 
\,.
\end{align}
\end{subequations}

\section{Gaussian and stationary phase approximations}
\label{sec/approximation-annex}

In this appendix, we give a more detailed analysis of the two approximation
schemes used in \cref{sec/excess} to understand the qualitative form of
the Wigner function of Gaussian wavepackets.

\subsection{Gaussian approximation}

For a general trajectory, computing analytically the Wigner function is
not possible. Nonetheless, meaningful information can be already 
uncovered by resorting to a Gaussian approximation. One has to 
approximate the wavefunction $\Phi_\alpha(\tau)$ has a Gaussian 
in time around its maximum value. From
\begin{equation}
	\Phi_\alpha(\tau) = 
	\sqrt{\frac{p_0}{\left( 2\pi \sigma_x^2 \right)^{1/2}}}
	\me^{-\left[ f_-(\tau)  + x_0 \right]^2/4\sigma_x^2}
	\,
	\me^{-\mi p_0\left[ f_-(\tau) + x_0 \right]}
\,,
\end{equation}
let's do an expansion around the maximum reached at time $\tau_r$:
$\int_0^{\tau_r} \exp(-A(u)) \, \md u = x_0$.
We now expand around this maximum. Thus, 
\begin{subequations}
\begin{align}
	f_-(\tau_r + \upsilon)
	&=
	x_0 + D_r \upsilon
	- \frac{1}{2} a_r D_r \upsilon^2
\\
	f_-(\tau_r + \upsilon)^2
	&=
	x_0^2 + 2 x_0 D_r \upsilon
	+ D_r(D_r - x_0 a_r) \upsilon^2
\end{align}
\end{subequations}
with $D_r = \me^{-A(\tau_r)}$ is the gravitational redshit shift, as we will see.
We can start by approximating the wavefunction as a Gaussian. We have
\begin{equation}
	\Phi(\tau_r + \upsilon)
	=
	\sqrt{\frac{p_0}{\left( 2\pi \sigma_x^2 \right)^{1/2}}}
	\exp\left(
		- \frac{D_r^2}{4 \sigma_x^2} \upsilon^2
		+ \frac{\mi}{2} p_0 a_r D_r \upsilon^2
		- \mi p_0 D_r \upsilon
	\right)
\,.
\end{equation}
We recognize here a Gaussian linear chirp, centered around frequency
$\omega_r = D_r p_0$, time $t_r$ and with a chirp rate $- \omega_r a_r$.
To this order, we find that the Wigner function associated with the
$\Phi^*\Phi$ and the $\Phi \Phi + \Phi^*\Phi^*$ terms are respectively 
given by 
\begin{subequations}
\begin{align}
W_{\Phi_\alpha\Phi_\alpha^*}(\tau,\omega) &= 
	\frac{2p_0}{D_r}
	\exp\left(
		-\frac{D_r^2}{2 \sigma_x^2} (\tau - \tau_r)^2
	\right)
	\exp\left(
		-\frac{1}{2}\frac{4\,\sigma_x^2}{D_r^2}
		\left(\vphantom{\rule{0pt}{2ex}}\omega - \omega_r(1-a_r (\tau-\tau_r))\right)^2
	\right)
	\label{eq/gaussian/approx/wf}
\\
W_{\Phi_\alpha\Phi_\alpha}(\tau,\omega) &= 
	\Re \frac{2}{\sqrt{
		D_r^2 - 2 \mi a_r \omega_r \sigma_x^2
	}}
	\exp\left(
		- \frac{D_r^2}{2 \sigma_x^2} (\tau-\tau_r)^2
		- \frac{1}{2} \frac{4\, \sigma_x^2}{D_r^2 + 4 a_r^2 \omega_r^2}
		\omega^2
	\right) \nonumber \\
	&\exp\left(
	\mi \omega_r a_r  \left(
			(\tau-\tau_r)^2
			+ \frac{\omega^2}{D_r^4/4\sigma_x^4 + a_r^2 \omega_r^2}
		\right)
	\right)
	\exp\left(2 \mi \omega_r (\tau-\tau_r)\right)
	\label{eq/gaussian/approx/wf-correlation}
\,.
\end{align}
\end{subequations}

\subsection{Stationary phase approximation}

The stationary phase approximation allows to write the approximate 
form:
\begin{align}
W(\tau,\omega) = 
	\sum_s \frac{(8\pi)^{1/2}}{|\partial^2_\upsilon \Phi(\tau_s;\tau,\omega)|^{1/2}}
	A(\tau_s ; \tau)
	\cos\left( \Phi(\tau_s ; \tau, \omega) + \frac{\pi}{4}\mathrm{sgn}\,\partial^2_\upsilon \Phi(\tau_s;\tau,\omega) \right)
\label{eq/Wigner-statphase}
\end{align}
On the points where the stationary phase is valid, the Wigner function 
(the term $\Phi_\alpha(t,\mathbf{x})\Phi_\alpha^*(t',\mathbf{x}') + \text{h.c.}$) 
of a Gaussian coherent state probed by a uniformly accelerated observer
is:
\begin{align}
W_{\Phi_\alpha\Phi_\alpha^*}(\tau,\omega) = 
	\frac{2p_0}{\sigma_x}
	\sqrt{\frac{2}{a\sqrt{\omega^2 - \omega^2(\tau)}}}
	&\exp\left(
		-\frac{1}{2(ap_0\sigma_x)^2}
		\left[
		(\omega-\omega_r)^2 + (\omega-\omega(\tau))(\omega+\omega(\tau))
		\right]
		\right) \nonumber \\
	&\cos\left(2a^{-1}\sqrt{\omega^2 - \omega^2(\tau)}- 
	2a^{-1}\omega \,\text{argcosh}\left(\frac{\omega}{\omega(\tau)}\right) + 
	\frac{\pi}{4} \right)
\end{align}
Since far away from the instantaneous frequency curve the Wigner 
function is decreasing in a Gaussian way, it is meaningful to expand the 
argument of the cosine function in $\omega/\omega(\tau)$. Then we have the 
approximate form 
\begin{align}
W_{\Phi_\alpha\Phi_\alpha^*}(\tau,\omega) = 
	\frac{2p_0}{\sigma_x}
	\sqrt{\frac{2}{a\sqrt{\omega^2 - \omega^2(\tau)}}}
	&\exp\left(
		-\frac{(\omega-\omega_r)^2 + [\omega-\omega(\tau)][\omega+\omega(\tau)]}{2(ap_0\sigma_x)^2}
		\right) \nonumber \\
	&\cos\left( \frac{2\omega}{a} \left[1-\ln2\omega/\omega(\tau) \right] +  \frac{\pi}{4}\right)
\,.
\end{align}
The logarithmic correction $\omega \ln \omega/p_0$ can also be neglected at first 
order. This gives the final approximate form
\begin{align}
W_{\Phi_\alpha\Phi_\alpha^*}(\tau,\omega) = 
	\frac{2p_0}{\sigma_x}
	\sqrt{\frac{2}{a\sqrt{\omega^2 - \omega^2(\tau)}}}
	&\exp\left(
		-\frac{(\omega-\omega_r)^2 + [\omega-\omega(\tau)][\omega+\omega(\tau)]}{2(ap_0\sigma_x)^2}
		\right) \nonumber \\
	&\cos\left( 2\left[1-\ln2\right] \frac{\omega}{a} - 2 \omega \tau + \frac{\pi}{4}\right)
\,.
\end{align}

\begin{figure}
	\includegraphics{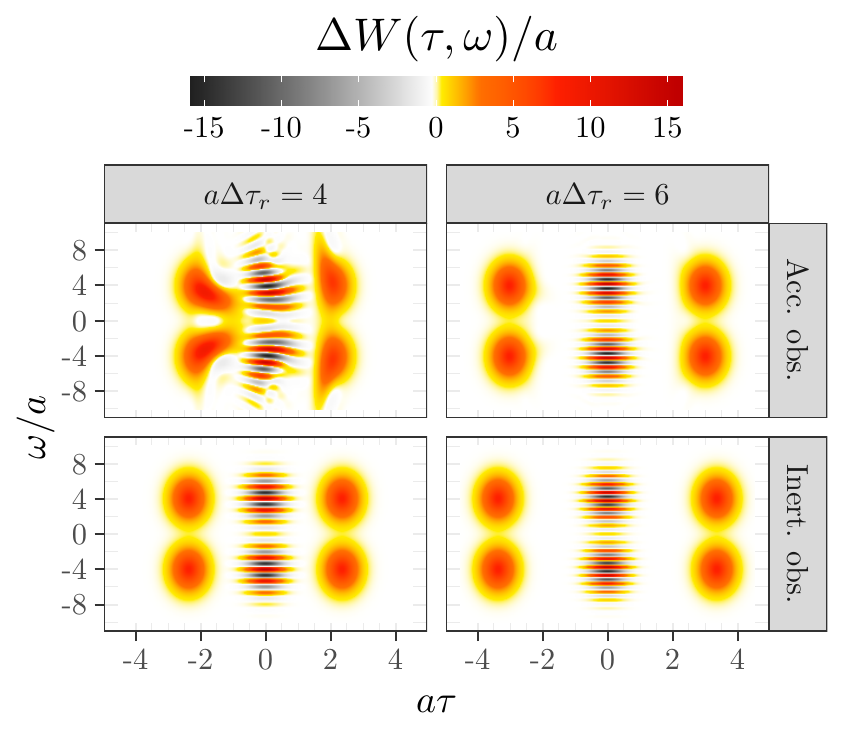}
	\caption{Comparison of the signal received from a superposition of coherent states
	for a uniformly accelerated (top row) and inertial detector (bottom row).
	$\Delta \tau_{r}$ is the time difference between the two members of
	the superposition in the accelerated frame. The time dilation for this twin
	paradox configuration is clear.
	(Parameters : Total time $a \tau_{\text{acc}} = 4$, transitions
	at $\tau= -2, -1, 1, 2$ (proper time), $p_0/a = 4$, $a \sigma_x = 1/3$.)
	\label{fig/comp-twin-super}}
\end{figure}